\documentclass[twocolumn,aps,unsortedaddress]{revtex4-1}

\usepackage{amsmath}
\usepackage{mathtools}
\usepackage{bm}
\usepackage{amssymb}
\usepackage[colorlinks=true,linkcolor=blue,citecolor=blue,urlcolor=blue]{hyperref}

\newcommand{\Tr}[1]{\mathrm{Tr}{\left\{#1\right\}}}
\newcommand{\rhoss}{\hat{\rho}_{\mathrm{ss}}}
\newcommand{\Ropt}[1]{{#1}^{\mathrm{R}}}
\newcommand{\Aopt}[1]{{#1}^{\mathrm{A}}}
\newcommand{\RAopt}[1]{{#1}^{\mathrm{R/A}}}

\DeclarePairedDelimiter\abs{\lvert}{\rvert}%
\DeclarePairedDelimiter\norm{\lVert}{\rVert}%

\makeatletter
\let\oldabs\abs
\def\abs{\@ifstar{\oldabs}{\oldabs*}}
\let\oldnorm\norm
\def\norm{\@ifstar{\oldnorm}{\oldnorm*}}
\makeatother

\begin{document}

\title{An Antisymmetric Berry Frictional Force At Equilibrium in the Presence of Spin-Orbit Coupling}

\author{Hung-Hsuan Teh}
\email{teh@sas.upenn.edu}
\affiliation{Department of Chemistry, University of Pennsylvania, Philadelphia, Pennsylvania 19104, USA}

\author{Wenjie Dou}
\email{douwenjie@westlake.edu.cn}
\affiliation{School of Science, Westlake University, Hangzhou, Zhejiang 310024, China}
\affiliation{Institute of Natural Sciences, Westlake Institute for Advanced Study, Hangzhou, Zhejiang 310024, China}

\author{Joseph E. Subotnik}
\email{subotnik@sas.upenn.edu}
\affiliation{Department of Chemistry, University of Pennsylvania, Philadelphia, Pennsylvania 19104, USA}

\date{\today}

\begin{abstract}
We analytically calculate the electronic friction tensor for a molecule near 
a metal surface in the case that the electronic Hamiltonian is complex-valued,
e.g. the case that there is spin-orbit coupling and/or an external magnetic field. In such a case, \textit{even at equilibrium},
we show that the friction tensor is not symmetric.  Instead, the tensor is the real-valued sum of one 
positive definite  tensor (corresponding to dissipation) plus one antisymmetric tensor
(corresponding to a Berry pseudomagnetic force). Moreover, we find
that this Berry force can be much larger than the dissipational force, suggesting
the possibility of strongly spin-polarized chemicurrents or strongly spin-dependent rate constants for systems with spin-orbit coupling.
\end{abstract}

\maketitle

\section{Introduction}\label{sec:intro}
When  nuclear degrees of freedom (DoF) are allowed to fluctuate in the presence of a continuum of electronic DoFs, not only will the electronic wave function change at different nuclear positions (in accordance with the Born-Oppenheimer picture) --- the nuclear wave packet will also change as it receives feedback from the electronic DoF.
In the adiabatic limit, the 
electronic feedback on the nuclei is composed of three parts: the adiabatic force $F_{\mu}$, the random force $\delta F_{\mu}$, and the frictional damping force. The friction tensor $\gamma_{\mu\nu}$ captures the strength of the nuclear damping force in the $\mu$-direction as caused by nuclear motion in the $\nu$ direction and reflects how fast electronic transitions inevitably interfere with simple nuclear (Newtonian) motion. Mathematically, the nuclei follow a stochastic Langevin equation\cite{van1992stochastic,dou2017born} of the form,
\begin{align}
M_{\mu}\ddot{R}_{\mu}=F_{\mu}-\sum_{\nu}\gamma_{\mu\nu}\dot{R}_{\nu}+\delta F_{\mu},\label{eq:langevin_eq}
\end{align}
where $M_{\mu}$ is the mass of a nuclei and $R_{\mu}$ is the nuclear position in the $\mu$ direction. 

Several important general properties (and proofs) about $\gamma_{\mu\nu}$, including the positive definiteness and the fluctuation-dissipation theorem, are provided in SM \ref{si:review_ft}-\ref{si:no_gamma_A_for_real_H_equil}. Historically, electronic friction was first considered as a first-order correction to the Born-Oppenheimer approximation for dynamics near a metal surface, and though there have been many separate approaches for calculating the electronic friction tensor\cite{suhl:1975:prl_elfriction, suhl:1975:prb_elfriction,tully:1995:electronic_friction,brandbyge:1995:elfriction,dundas:2012:prb_berry,persson:2005:friction_noncondon, persson:2007:friction_noncondon,hynes:1993:elfriction,langreth:1999:prb_elfriction,langreth:1998:prb_elfriction,mozyrsky:2006:elfriction,mozyrsky:2007:pre_friction,lu2010blowing,bode2012current,chen2018current,dou2017born,galperin:2015:friction,dou2018perspective,lu2019semi}, all agree in the Markovian limit\cite{dou2017born}.
$\gamma_{\mu\nu}$ can be divided into a symmetric part $\gamma_{\mu\nu}^{\mathrm{S}}$ and an antisymmetric part $\gamma_{\mu\nu}^{\mathrm{A}}$. For a system at equilibrium, $\gamma_{\mu\nu}^{\mathrm{S}}$ is positive definite; this term can only dissipate energy to the surroundings (and avoid unstable dynamics). Such relaxation processes have been reported as important for molecule-metal interface dynamics (scattering, adsorption etc.)\cite{huang2000observation,huang2000vibrational,bunermann2015electron,rittmeyer2018energy}, electron transfer within electronic devices\cite{pop2010energy}, and heating due to the phonon motion\cite{koch2004thermopower,kaasbjerg2013charge}.  The diagonal component ($\mu=\nu$) can significantly change the electron-hole pair induced vibrational lifetime\cite{juaristi:2008:prl_ldfa,reuter:2015:prl_ldfa}, and the off-diagonal elements of a symmetric friction tensor ($\mu\neq\nu$)  can also be crucial\cite{askerka2016role,maurer2016ab}.

We have far less experience with the antisymmetric component of the friction tensor, $\gamma_{\mu\nu}^{\mathrm{A}}$, which contributes a Lorentz-like force. 
Within the chemical physics condensed matter community, the usual assumption is that $\gamma_{\mu\nu}^{\mathrm{A}} = 0$ at equilibrium. That being said, for a strictly real-valued Hamiltonian describing a typical molecule on a typical metal, von Oppen and others\cite{lu2010blowing,bode2012current} have demonstrated that $\gamma_{\mu\nu}^{\mathrm{A}}\ne0$ only when molecules are in contact with two metals that are {\em out of} equilibrium (i.e. with a current).

Now, within the description above\cite{lu2010blowing,bode2012current}, there has been the assumption of a strictly real-valued  Hamiltonian. However, for molecule-metal interfaces or for surface heterostructures, due to the short electron screening length of a metal, the effective electric field gradient on the surface should lead to strong Rashba spin-orbit coupling\cite{manchon2015new}. Furthermore, a built-in molecular spin-orbit coupling can be enhanced due to molecular geometry, i.e. molecules with large curvature or torsion in geometry are believed to have larger spin-orbit coupling\cite{ando2000spin,huertas2006spin}. For these reasons,
{\em a complex-valued Hamiltonian may be quite relevant.} Moreover, Robbins and Berry have demonstrated that antisymmetric (Berry) forces may appear if the Hamiltonian is complex-valued\cite{berry1993chaotic}. Exact scattering calculations have shown that, for a closed model complex-valued Hamiltonian, the resulting Berry force effects can be large and strongly  affect electron transfer processes\cite{miao2019extension}.

Where does this leave us as far as understanding molecular dynamics near a metal surface? For the most part, a molecule on a metal surface is occupied by a fractional number of electrons; as electrons are shared between molecule and metal, the molecular nuclei will feel different forces depending on the fluctuating electronic charge state.  While such fluctuations are known to lead to fast vibrational relaxation of a diatomic on a metal surface\cite{huang2000vibrational} through the symmetric tensor, one can ask: Does a significant nonzero antisymmetric friction tensor (i.e. a pseudomagnetic field) also appear when we consider a complex-valued Hamiltonian describing a molecule near a metal surface in equilibrium?
Note that, except for a few analogous examples in the realm of spintronics\cite{bajpai2020spintronics}, to date, the effect of a magnetic field or spin-orbit coupling has been ignored in friction tensor calculations (even though $\gamma_{\mu\nu}$ is general). Note also that Ref. \citenum{wu2021electronic} predicts that  a huge Berry force can be generated for an isolated molecular system near a sharp avoided crossing in the presence of spin-orbit coupling.  
One must wonder whether such a huge Berry force will still exist when one considers a molecule strongly coupled to a metal surface.

Below, we will address these questions. In particular, we will show that: (i) Like the case of an isolated molecular system, a Berry force exists whenever a molecular system with a complex-valued Hamiltonian is coupled to a bath (no matter whether or not the total system is in \textit{equilibrium}). (ii) Unlike the case of an isolated molecular system, the strength of the Berry force does not require a tiny energy gap (i.e. a sharp avoided crossing) to achieve a large Berry force (in fact an energy gap is necessary). (iii) $\gamma_{\alpha\nu}^{\mathrm{A}}$ is comparable or can even be one order of magnitude larger than $\gamma_{\mu\nu}^{\mathrm{S}}$ and thus affects the experimental observable. These facts suggest that nuclear motion at surfaces should promote certain flavor of electronic spin selectivity, e.g. perhaps spin selectivity in transport with chiral molecules.


\section{Model System}\label{sec:model_sys}
We consider a model in which a two-level system is coupled to two leads and the two-level system depends on two dimensional nuclear DoF. While there is an immense amount known about the (symmetric) friction tensor that arises for a resonant level model\cite{bode2012current,chen2019electronic}, no such results or intuition have been derived for the antisymmetric friction tensor even in the case of a two-level model at equilibrium.

Here we will derive the friction tensor analytically. The total electronic Hamiltonian $\hat{H}$ is divided into three components, the system $\hat{H}_{\mathrm{s}}$, the bath $\hat{H}_{\mathrm{b}}$ and the system-bath coupling $\hat{H}_{\mathrm{c}}$:
\begin{align*}
\hat{H}=&\hat{H}_{\mathrm{s}}+\hat{H}_{\mathrm{b}}+\hat{H}_{\mathrm{sb}},\\
\hat{H}_{\mathrm{s}}=&\sum_{mn}h^{\mathrm{s}}_{mn}(\mathbf{R})\hat{b}_{m}^{\dagger}\hat{b}_{n}+U(\mathbf{R}),\\
\hat{H}_{\mathrm{b}}=&\sum_{k\alpha}\epsilon_{k\alpha}\hat{c}_{k\alpha}^{\dagger}\hat{c}_{k\alpha},\\
\hat{H}_{\mathrm{c}}=&\sum_{m,k\alpha}V_{m,k\alpha}\hat{b}_{m}^{\dagger}\hat{c}_{k\alpha}+\mathrm{H.c.}
\end{align*}
Above, $m$, $n$ label system orbitals, and $\hat{b}_{m}^{\dagger}$ ($\hat{b}_{m}$) creates (annihilates) an electron in the system orbital $m$. $\hat{c}_{k\alpha}^{\dagger}$ ($\hat{c}_{k\alpha}$) creates (annihilates) an electron in the $k$-th orbital of a lead $\alpha$.  Note that, for the sake of generality, 
all expressions below will be derived for the case of two electronic leads; $\alpha=\mathrm{L},\mathrm{R}$ indicates left and right leads.  If we set the fermi levels $\mu_{\alpha}$ of both leads to be equal, $\mu_{\mathrm{L}} = \mu_{\mathrm{R}}$, there is no difference between having
one lead (with $2N$ orbitals) or two leads (with $N$ orbitals).
$h^{\mathrm{s}}_{mn}$ is the molecular electronic  Hamiltonian  that depends explicitly on $\mathbf{R}$, the molecular nuclear DoFs,
and we know that this dependence on $\mathbf{R}$  leads to a symmetric friction tensor\cite{dou2017born}. $U(\mathbf{R})$ is a pure nuclear potential energy. $V_{m,k\alpha}$ represents the tunneling element between the system orbital $m$ and the lead orbital $k\alpha$,
which we assume independent of $\mathbf{R}$ (the so-called Condon approximation). Within this model, the most general system Hamiltonian can be written in Pauli matrices representation ($\sigma_{i}$\cite{pauli_representations}) as:
\begin{align*}
\mathbf{h}^{\mathrm{s}}=\mathbf{h}(x,y)\cdot\bm{\sigma}=\sum_{i=1,2,3}h_{i}(x,y)\sigma_{i},
\end{align*}
where $\lbrace h_{i}\rbrace$ is real. Note that the inclusion of $h_{2}$ makes the Hamiltonian possibly complex-valued, as might arise from an external magnetic field or spin-orbit coupling.
 
If we now evaluate $\gamma_{\mu\nu}$ for the case of non-interacting electrons (following Ref. \citenum{dou2018universality}; $\hbar=1$), we find that the (exact) final results is (see SM \ref{si:review_ft}):
\begin{align}
\gamma_{\mu\nu}=\int\frac{d\epsilon}{2\pi}\Tr{\partial_{\mu}h^{\mathrm{s}}\partial_{\epsilon}\Ropt{G}\partial_{\nu}h^{\mathrm{s}}G^{<}}+\mathrm{H.c.},\label{eq:ft_non_condon}
\end{align}
where $\Ropt{G}=(\epsilon-h^{\mathrm{s}}-\Ropt{\Sigma})^{-1}$ is the (two-level) system retarded Green's function, $\Ropt{\Sigma}_{mn}=\sum_{k\alpha}V_{m,k\alpha}\Ropt{g}_{k\alpha}V^{*}_{n,k\alpha}$ is the system self energy, and $\Ropt{g}_{k\alpha}=(\epsilon-\epsilon_{k\alpha}+i\eta)^{-1}$ is the lead retarded self energy ($\eta\rightarrow0^{+}$). $G^{<}$ is the system lesser Green's function and, provided that an imaginary surrounding is quadratic\cite{dou2018universality} or the system spectral broadening due to the leads is finite\cite{haug2008quantum}, $G^{<}$ can be calculated by the Keldysh equation, $G^{<}=\Ropt{G}\Sigma^{<}\Aopt{G}$. Here $\Sigma^{<}_{mn}=\sum_{k\alpha}V_{m,k\alpha}g^{<}_{k\alpha}V_{n,k\alpha}^{*}$ is the system lesser self energy, and $g^{<}_{k\alpha}(\epsilon)=i2\pi f_{\alpha}(\epsilon)\delta(\epsilon-\epsilon_{k\alpha})$ is the lead lesser Green's function. ($f(\epsilon)=1/[\exp{(\beta(\epsilon-\mu))}+1]$ is the Fermi-Dirac distribution with the inverse temperature $\beta$ and the chemical potential $\mu$.)

We further make the following standard assumptions\cite{stefanucci2013nonequilibrium}: (i) the tunneling-width matrix $\Gamma_{mn}=2\pi\sum_{k\alpha}V_{m,k\alpha}V_{n,k\alpha}^{*}\delta(\epsilon-\epsilon_{k\alpha})$ is independent of $\epsilon$ (i.e. the wide band limit approximation), (ii) $\Gamma_{mn}=\tilde{\Gamma}$ is a constant, (iii) the tunneling elements $V_{m,k\alpha}$ are independent of $k$, and (iv) the left lead couples only to orbital $1$ and the right lead couples only to orbital $2$, with the two coupling constants the same real value. Therefore,
$\Ropt{\Sigma}=-\frac{i}{2}\tilde{\Gamma}$, and $\Sigma^{<}=i\tilde{\Gamma}\left(f_{\mathrm{L}}\,0;\,0\,f_{\mathrm{R}}\right)$ where $f_{\mathrm{L}}$ and $f_{\mathrm{R}}$ are Fermi-Dirac distribution of the left and right leads respectively. Please see SM \ref{si:schematic_pic} for a schematic picture (equilibrium case).

A calculation (see SM \ref{si:ft_calculation_details}) shows that the friction tensor (from Eq. (\ref{eq:ft_non_condon})) is,
\begin{alignat}{2}
\gamma_{\mu\nu}=&\gamma_{\mu\nu}^{\mathrm{S}}+\gamma_{\mu\nu}^{\mathrm{A}},\label{eq:ft_tls}\\
\gamma_{\mu\nu}^{\mathrm{S}}=&\frac{2}{\pi}\int_{-\infty}^{\infty}d\epsilon\bigg\{&&-2\mathfrak{Re}\left\{C\tilde{\epsilon}\right\}\left(\partial_{\mu}\mathbf{h}\cdot\partial_{\nu}\mathbf{h}\right)\left(\mathbf{h}\cdot\bm{\kappa}\right)\notag\\
&&&+2\mathfrak{Re}\left\{C\tilde{\epsilon}\right\}\left(\partial_{\mu}\mathbf{h}\cdot\mathbf{h}\right)\left(\partial_{\nu}\mathbf{h}\cdot\bm{\kappa}\right)\notag\\
&&&+2\mathfrak{Re}\left\{C\tilde{\epsilon}\right\}\left(\partial_{\nu}\mathbf{h}\cdot\mathbf{h}\right)\left(\partial_{\mu}\mathbf{h}\cdot\bm{\kappa}\right)\notag\\
&&&+\kappa_{0}\mathfrak{Re}\left\{C\left(\tilde{\epsilon}^{2}+h^{2}\right)\right\}\partial_{\mu}\mathbf{h}\cdot\partial_{\nu}\mathbf{h}\bigg\}\label{eq:ft_tls_sym}\\
\gamma_{\mu\nu}^{\mathrm{A}}=&\frac{2}{\pi}\int_{-\infty}^{\infty}d\epsilon\bigg\{&&-\mathfrak{Im}\left\{C\left(\tilde{\epsilon}^{2}+h^{2}\right)\right\}
\bm{\kappa}\cdot\left(\partial_{\mu}\mathbf{h}\times\partial_{\nu}\mathbf{h}\right)\notag\\
&&&+2\kappa_{0}\mathfrak{Im}\left\{C\tilde{\epsilon}\right\}\mathbf{h}\cdot\left(\partial_{\mu}\mathbf{h}\times\partial_{\nu}\mathbf{h}\right)\bigg\},\label{eq:ft_tls_antisym}
\end{alignat}
Here
\begin{align*}
C\equiv-\left(\frac{1}{\tilde{\epsilon}^{2}-h^{2}}\right)^{2}i\tilde{\Gamma}\abs{\frac{1}{\tilde{\epsilon}^{2}-h^{2}}}^{2},
\end{align*}
where $\tilde{\epsilon}=\epsilon+i\tilde{\Gamma}/2$ is a complex number, and $h^{2}=\mathbf{h}\cdot\mathbf{h}$. The $\kappa$'s are defined as
\begin{align*}
\kappa_{0}=&\frac{1}{2}\left[\left(f_{\mathrm{L}}+f_{\mathrm{R}}\right)\left(h_{1}^{2}+h_{2}^{2}\right)+f_{\mathrm{L}}\abs{\tilde{\epsilon}+h_{3}}^{2}+f_{\mathrm{R}}\abs{\tilde{\epsilon}-h_{3}}^{2}\right],\\
\kappa_{1}=&\mathfrak{Re}\left\{\left[f_{\mathrm{L}}\left(\tilde{\epsilon}^{*}+h_{3}\right)+f_{\mathrm{R}}\left(\tilde{\epsilon}-h_{3}\right)\right]\left(h_{1}+ih_{2}\right)\right\},\\
\kappa_{2}=&\mathfrak{Im}\left\{\left[f_{\mathrm{L}}\left(\tilde{\epsilon}^{*}+h_{3}\right)+f_{\mathrm{R}}\left(\tilde{\epsilon}-h_{3}\right)\right]\left(h_{1}+ih_{2}\right)\right\},\\
\kappa_{3}=&\frac{1}{2}\left[\left(f_{\mathrm{R}}-f_{\mathrm{L}}\right)\left(h_{1}^{2}+h_{2}^{2}\right)+f_{\mathrm{L}}\abs{\tilde{\epsilon}+h_{3}}^{2}-f_{\mathrm{R}}\abs{\tilde{\epsilon}-h_{3}}^{2}\right].
\end{align*}
Note that $\kappa_{0}$ and $\bm{\kappa}=(\kappa_{1},\kappa_{2},\kappa_{3})$ are real functions. When the total system is in equilibrium, namely $f_{\mathrm{L}}=f_{\mathrm{R}}=f$,
\begin{align}
\kappa_{0}=&f\left(\epsilon^{2}+h^{2}+\frac{\Gamma^{2}}{4}\right),\label{eq:condition1_at_equil}\\
\bm{\kappa}=&2f\epsilon\mathbf{h}.\label{eq:condition2_at_equil}
\end{align}
Equations (\ref{eq:ft_tls})-(\ref{eq:ft_tls_antisym}) represent a very general electronic friction tensor for a minimal model of a molecule near metal surfaces in the presence of spin-orbit coupling.

\section{Results and Discussions}\label{sec:results_n_discussions}
According to Eq. (\ref{eq:ft_tls}) we can make two important and general conclusions regarding Berry forces of equilibrium near a metal surface.
First, according to Eqs (\ref{eq:ft_tls_antisym})-(\ref{eq:condition1_at_equil}), $\gamma_{\mu\nu}^{\mathrm{A}}$ is proportional to $\mathbf{h}\cdot\left(\partial_{\mu}\mathbf{h}\times\partial_{\nu}\mathbf{h}\right)$. Therefore, $\gamma_{\mu\nu}^{\mathrm{A}}$ will vanish when at least one element of $\mathbf{h}$ is zero, or when two elements of $\mathbf{h}$ are the same. These facts demonstrate not only that an imaginary off-diagonal coupling ($h_{2}$) is required for a nonzero $\gamma_{\mu\nu}^{\mathrm{A}}$, but also that the key source of a nonzero $\gamma_{\mu\nu}^{\mathrm{A}}$ is the spatial dependence of the phase of the off-diagonal coupling, $\tan^{-1}{(h_{2}/h_{1})}$. After all, if $h_{1}=0$ or $h_{1}(x,y)=h_{2}(x,y)$, we can find a constant change of basis transformation
that guarantees a globally real-valued Hamiltonian and therefore $\gamma_{\mu\nu}^{\mathrm{A}}=0$. In other words, in such a case, there is no Lorentz-like force.

Second, according to Eqs. (\ref{eq:ft_tls})-(\ref{eq:condition2_at_equil}), one can construct several nonequivalent Hamiltonians that generate equivalent friction tensors. To see this, note that, when the system is in equilibrium, the symmetric terms in Eq. (\ref{eq:ft_tls}) all have dot product dependence on $\mathbf{h}$, namely $h^{2}$, $\sum_{i}\partial_{\mu}h_{i}\partial_{\nu}h_{i}$ and $\sum_{i}\partial_{\mu}h_{i}h_{i}$. Thus, the symmetric terms are invariant to any permutation of $\mathbf{h}=\{h_{1},h_{2},h_{3}\}$. Moreover, the two terms comprising $\gamma_{\mu\nu}^{\mathrm{A}}$ depend on $\mathbf{h}\cdot(\partial_{\mu}\mathbf{h}\times\partial_{\nu}\mathbf{h})$, which are also invariant under cyclic permutation of the $\mathbf{h}$ elements. Thus, different Hamiltonians can generate the same friction tensor and, as a practical matter, this must have experimental consequences as some Hamiltonians are undoubtedly easier to realize than others. For example, in Eq. (\ref{eq:hs_ft}) we will consider a model Hamiltonian with diagonal coupling $h_{3}=x+\Delta$; here, as in standard Marcus theory, $\Delta$ is a driving force that will be shown to play an important role in generating a
large antisymmetric friction tensor. Nevertheless, if ones imagines permuting the $\mathbf{h}$ elements by substituting $h_{1}\rightarrow h_{3}\rightarrow h_{2}$, then the parameter $\Delta$ will enter on the off-diagonal of the Hamiltonian and can be realized, e.g., by tuning an external magnetic field.

These are the only direct, general conclusions we can make from Eqs. (\ref{eq:ft_tls})-(\ref{eq:condition2_at_equil}). Next, let us focus on a model problem which can yield further insight using numerical analysis. We imagine the standard case of two shifted parabolas, expressed in a nuclear space with two dimensions and with a driving force of $2\Delta$. Mathematically, the system Hamiltonian is taken to be of the form:
\begin{align}
\mathbf{h}^{\mathrm{s}}=
\begin{pmatrix}
x+\Delta & Ax-iBy\\
Ax+iBy & -x-\Delta
\end{pmatrix},\label{eq:hs_ft}
\end{align}
and $U=x^{2}/2+y^{2}/2+1/2$. We calculate the electronic friction tensor by using Eq. (\ref{eq:ft_tls}). Note that the pure nuclear potential $U$ does not contribute to the friction tensor. Recall that $\gamma_{\mu\nu}^{\mathrm{A}}\propto\mathbf{h}^{\mathrm{s}}\cdot(\partial_{\mu}\mathbf{h}^{\mathrm{s}}\times\partial_{\nu}\mathbf{h}^{\mathrm{s}})=AB\Delta$.  Thus, as argued above, if there is no change in the phase of the off-diagonal coupling ($A=0$ or $B=0$) in the nuclear space, we will find that $\gamma_{xy}^{\mathrm{A}}=0$. Also notice that when the driving force $\Delta=0$, again $\gamma_{\mu\nu}^{\mathrm{A}}=0$.  Beyond these two extreme cases, we will find both symmetric and antisymmetric components of the friction tensor.

\begin{figure}[!h]
\centering
\includegraphics[width=.5\textwidth]{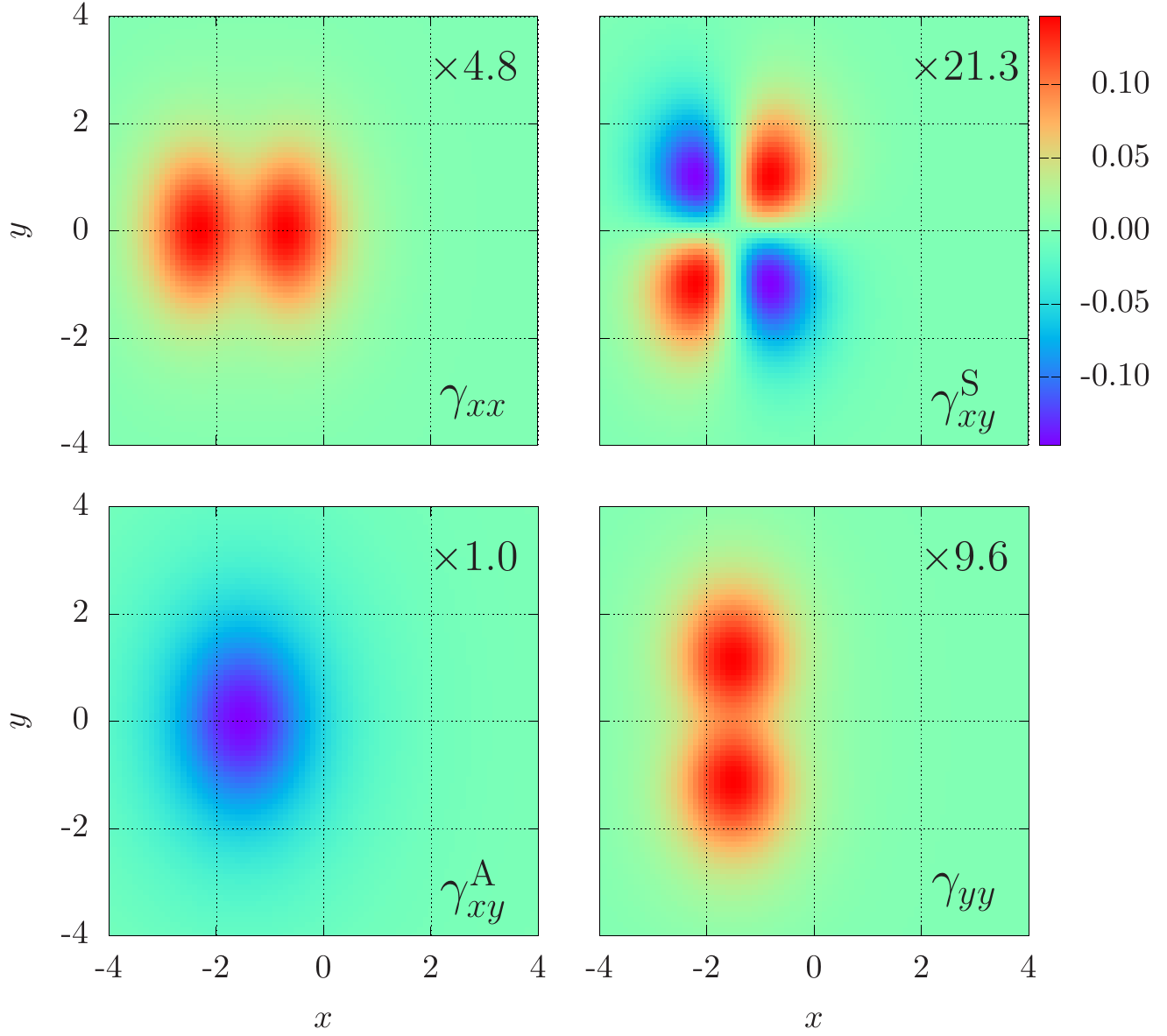}
\caption{Friction tensor calculation results: $\gamma_{xx}$ (top left), $\gamma_{xy}^{\mathrm{S}}$ (top right), $\gamma_{xy}^{\mathrm{A}}$ (bottom left) and $\gamma_{yy}$ (bottom right). Parameters: $\tilde{\Gamma}=1$, $\mu_{\mathrm{R}}=\mu_{\mathrm{L}}=0$, $\beta=2$, $A=1$, $B=1$, $\Delta=3$. Notice that all the results in Figs. \ref{fig:D3} and \ref{fig:b2_D3} have mirror symmetry about $x=-1.5=-\Delta/(A^{2}+1)$ and $y=0$, because all six terms in Eq. (\ref{eq:ft_tls_sym}) and Eq. (\ref{eq:ft_tls_antisym}) are functions of $[x+\Delta/(A^{2}+1)]^{2}$ and $B^{2}y^{2}$ when the system is in equilibrium.\label{fig:b2_D3}}
\end{figure}

In Fig. \ref{fig:b2_D3}, we show contour plots for the friction tensor with $\beta=2$, $A=B=1$. Here, $A=B$ corresponds to a strong change of phase in the off-diagonal coupling. Several features are clear from the contour plot. First, the antisymmetric friction tensor $\gamma_{xy}^{\mathrm{A}}$ is one order larger than all other symmetric friction tensors. Thus, clearly Lorentz-like motion can be as important as any dissipative process. Second, the magnitude of $\gamma_{xy}^{\mathrm{A}}$ is maximized around the avoided crossing at $(-1.5,0)$, but for each component of the symmetric friction tensor, the magnitude is maximized far from the avoided crossing. Therefore, depending on the preparation of the initial nuclear wave packet, one might imagine that slow nuclei will equilibrate before feeling any Lorentz like force. That being said, the exact details for any calculation must be evaluated on a case-by-case basis.

\begin{figure}[!h]
\centering
\includegraphics[width=.5\textwidth]{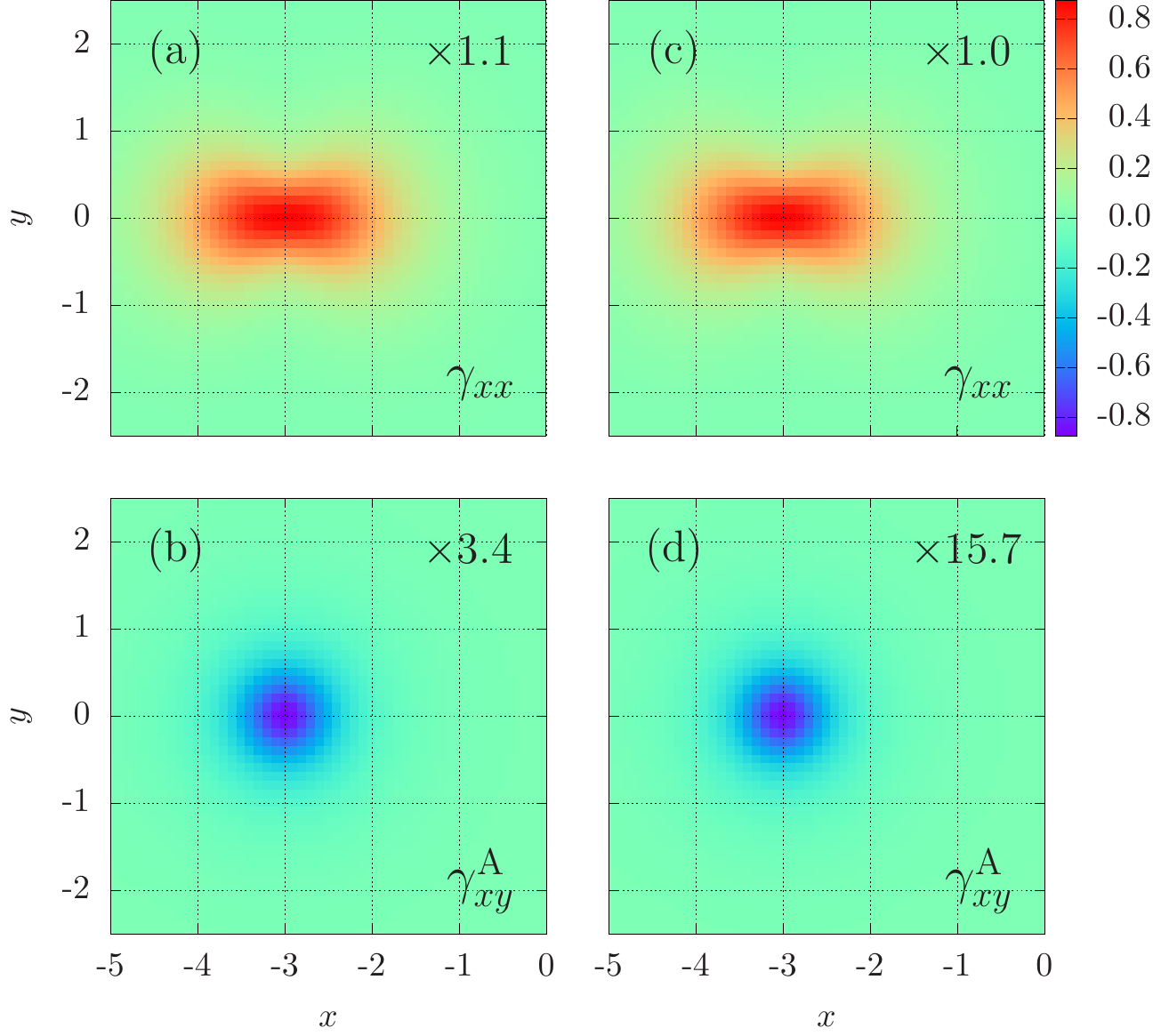}
\caption{Friction tensor calculation results (only $\gamma_{xx}$ and $\gamma_{xy}^{\mathrm{A}}$ are shown). Parameters for (a) and (b) are $\tilde{\Gamma}=1$, $\mu_{\mathrm{R}}=\mu_{\mathrm{L}}=0$, $\beta=2$, $A=0.05$, $B=1$, $\Delta=3$; (c) and (d) have the same parameters as (a) and (b), except that $A=0.01$.\label{fig:b2_D3_A005}}
\end{figure}

In Fig. \ref{fig:b2_D3_A005}, we further investigate how the relative strength of the antisymmetric friction tensor changes as a function of
how the off-diagonal coupling changes phase. Here we keep the same parameters as in Fig. \ref{fig:b2_D3}, except we change $A$. In subplots (a)(b), the antisymmetric friction tensor $\gamma_{xy}^{\mathrm{A}}$ has the same order of magnitude as the symmetric friction tensors. In subplots (c)(d), we reduce $A$ so that $A=0.01$; we find the antisymmetric friction tensor approaches zero rapidly. We conclude that in \textit{any} theoretical or experimental works which consider an external magnetic field or/and spin-orbit coupling with reasonable changes for the phase (here larger than $5\%$) in the nuclear space, we cannot ignore the effect of the antisymmetric friction tensor. We have also performed several \textit{ab initio} calculations so as to extract parameters for a real system --- a diphenylmethane junction (See SM \ref{si:diphenylmethane}). The results are consistent with the conclusion above --- even when spin-orbit coupling is small, $\gamma_{xy}^{\mathrm{A}}$ can still be dominant.


Lastly, before concluding, we summarize a few results that are addressed
in the SM. First, we investigate the dependence of $\gamma^{\mathrm{S}}$ and $\gamma_{\mathrm{A}}$ on $\beta$ and $\tilde{\Gamma}$ in SM \ref{si:ft_T_effect} and SM \ref{si:ft_other_results} respectively. We find that the relative strength of $\gamma_{xy}^{\mathrm{A}}$ grows stronger for lower temperatures. Also, when the system-bath coupling strength $\tilde{\Gamma}$ grows very large, both the symmetric and antisymmetric friction tensors become smaller and these tensors are nonzero over an effectively smaller portion of nuclear configuration space. Second, while we have considered an avoided crossing above, in SM \ref{si:ci} we investigate true complex-valued conical intersections. There, we show that the dynamical effect of a pseudomagnetic field in the direct vicinity of a true conical intersection is likely not very large. Third, and most importantly, throughout this letter, we have focused mostly on the magnitude of the antisymmetric friction tensor. Note that, in a basis of spin orbitals, switching spin up and spin 
down orbital will swap $h_{2}$ and $-h_{2}$ and lead to different signs of $\gamma_{\mu\nu}^{\mathrm{A}}$. Thus, different spins will feel different directions of the Lorentz force and the present formalism may underlie spin selectivity for molecular processes near metal surfaces\cite{naaman:2011:science:ciss_dna,naaman:2019:natrev,zollner2020insight}. As a practical matter if we were to construct an overall electronic friction tensor for the case of a system with multiple spin degrees of freedom, such a tensor would be meaningful only when the spin degrees of freedom interconvert rapidly, so that nuclear motion remains the slowest process of interest; alternatively, one would require separate friction tensors (one for up spin, and one for down spin) as in the present letter.

\section{Conclusions and Outlook}\label{sec:conclusions}
We have demonstrated that a large Lorentz force can operate on nuclei  \textit{in equilibrium} for systems with \textit{complex-valued} Hamiltonians. For a simple model of  two shifted parabolas, with spin-orbit coupling, according to an analytic expression for the friction tensor, the magnitudes of the relevant frictional components ($\gamma_{\mu\nu}^{\mathrm{S}}$ and $\gamma_{\mu\nu}^{\mathrm{A}}$) can be controlled by tuning the driving force $\Delta$ and the inverse temperature $\beta$. The antisymmetric part can be one order larger than the symmetric part for low temperatures. Moreover, $\gamma_{\mu\nu}^{\mathrm{A}}$ and $\gamma_{\mu\nu}^{\mathrm{S}}$ can be of comparable
magnitude even when the phase change of the off-diagonal coupling is very small. All of these results show that,
 for any relaxation processes with an external magnetic field or/and spin-orbit coupling, careful consideration of a Lorentz force due to the nuclear Berry curvature is necessary. We have also provided \textit{ab initio} calculations of a diphenylmethane junction, showing the same conclusions as above.

Looking forward, one can imagine two scenarios whereby the antisymmetric nature of $\gamma_{\mu\nu}$ will be paramount. First, if one scatters a molecule off a surface in the presence of spin-orbit coupling
, there is the real possibility that the presence of $\gamma_{\mu\nu}^{\mathrm{A}}$ will contribute meaningfully to a spin-polarized chemicurrent. Second, there is a deep question about whether the Lorentz force described here can help explain spin-selectivity as found in chiral-induced spin selectivity (CISS) experiments\cite{naaman:2011:science:ciss_dna,naaman:2019:natrev,zollner2020insight}. In other words, if nuclear wave packets attached to different spins feel different forces, might not one consequence of such a difference be spin-polarized of a current through a system where nuclei and electrons interact? Such a claim might be verified experimentally by the presence of an isotopic CISS effect. 
Finally, from a theoretical point of view, note that a recent
 paper has argued empirically that, for a molecule in the gas phase,  the Lorentz force is accentuated
dynamically when the molecule passes near a conical intersection that
is slightly modified by spin-orbit coupling\cite{wu2021electronic}.  
For our part, we find a similar result near a metal surface, i.e. the effect of the Lorentz force is maximized 
if $\hat{H}_\mathrm{s}$ admits an energy gap rather than displaying a true conical intersection.
In fact, according to Eq. (\ref{eq:ft_tls_antisym}), the antisymmetric part of the  friction tensor is zero
if one considers a gapless two-dimensional linear vibronic complex-valued Hamiltonian (See SM \ref{si:ci})\cite{ci_3d_note}.


Finally, in this letter we have analyzed the antisymmetric friction tensor that appears at equilibrium (with spin-orbit coupling). At the same time, a set of recent calculations has shown that a Berry force also appears when a molecule is placed between two leads out of equilibrium (without spin-orbit coupling)\cite{lu2010blowing,bode2012current}. To that end, the question remains as to what is the relationship between the equilibrium Lorentz force
analyzed here (derived in the case of a complex-valued Hamiltonian) and the previously published nonequilibrium Lorentz forces (derived in the case of real-valued Hamiltonian)\cite{dundas:2012:prb_berry,bode2012current}.  One can ask: Can the two Lorentz forces add to each other constructively? Can the forces be controlled individually by the properties of two leads? The present approach opens up the door to merge spintronics and nonadiabatic dynamics for an accurate description of spin-dependent current induced forces.

\bibliography{main}

\begin{thebibliography}{55}
\bibitem [{pau()}]{pauli_representations}%
  \BibitemOpen
  \href@noop {} {}\bibinfo {note} {$\sigma_{1}=\sigma_{x}$,
  $\sigma_{2}=\sigma_{y}$ and $\sigma_{3}=\sigma_{z}$. We hold $x$ and $y$ for
  the nuclear coordinates.}\BibitemShut {Stop}%
\bibitem [{ci_()}]{ci_3d_note}%
  \BibitemOpen
  \href@noop {} {}\bibinfo {note} {One must wonder whether such a result will
  still hold if we consider three-dimensional models of conical intersections,
  where the conical intersection point can map to a legitimate 3D branching
  plane. At present, we do not have a definitive answer to this
  question.}\BibitemShut {Stop}%
\end{thebibliography}

\onecolumngrid
\newpage

\renewcommand{\thesection}{\Alph{section}}
\renewcommand{\theequation}{\thesection\arabic{equation}}
\renewcommand{\thefigure}{\thesection\arabic{figure}}
\setcounter{section}{0}
\setcounter{equation}{0}
\setcounter{figure}{0}

\centerline{\bf\large Supplementary Material}

In this SM, we provide a host of details as relevant to the main findings of the letter. 
In sections \ref{si:review_ft}-\ref{si:no_gamma_A_for_real_H_equil}, we both review and prove several properties of the friction tensor based on the quantum-classical Liouville equation.  For instance (from Eq. (\ref{eq:general_ft}) below), it follows  that $\gamma_{\mu\nu}^{*}=\gamma_{\mu\nu}$, so that the friction
tensor is real-valued. Beyond this obvious statement, in Sec. \ref{si:review_ft}, we show that the friction tensor can be dramatically simplified in the absence of electron-electron interactions (leading to Eq. (\ref{eq:ft_non_condon}) in the main text of the letter). In Sec. \ref{si:positive_definite_ft_equil}, we prove that the friction tensor is positive definite at equilibrium (with or without interactions). In Sec. \ref{si:fd_theorem_based_on_qcle}, we prove that the fluctuation-dissipation theorem holds (though we note that a previously reported proof was incorrect). In Sec. \ref{si:no_gamma_A_for_real_H_equil}, we prove that, for a real-valued Hamiltonian, the friction tensor is purely symmetric.  Armed with these facts, in section \ref{si:ft_calculation_details} of this SM, we derive Eqs. (\ref{eq:ft_tls})-(\ref{eq:ft_tls_antisym}) of the main text of the letter. In sections \ref{si:ci}-\ref{si:ft_other_results}, we analyze friction in three different scenarios: \ref{si:ci}) the presence of a conical intersection, \ref{si:ft_T_effect}) a changing temperature,   and \ref{si:ft_other_results}) a variable coupling to the environment. In Sec. \ref{si:diphenylmethane}, we apply our findings to a realistic molecule, diphenylmethane, between two gold atoms. We show that, for a realistic set of parameters, the antisymmetric friction tensor can be large --- even for small spin-orbit couplings.  

\section{The Non-Interacting Friction Tensor Under the Condon Approximation}\label{si:review_ft}
In Ref. \citenum{dou2017born}, based on the quantum-classical Liouville equation (QCLE) and appropriate usages of the adiabatic theorem, a universal Fokker-Planck equation for a real-valued or complex-valued Hamiltonian was derived, either at equilibrium or in a nonequilibrium steady state with a Markovian\cite{galperin_note,galperin:2015:friction,dou2018perspective,lu2019semi} electronic friction tensor of the form:
\begin{align}
\gamma_{\mu\nu}=-\int_{0}^{\infty}dt\,\Tr{\partial_{\mu}\hat{H}e^{-i\hat{H}t}\partial_{\nu}\rhoss e^{i\hat{H}t}}.\label{eq:general_ft}
\end{align}
Here $\hat{H}$ is the electronic Hamiltonian and $\rhoss(\mathbf{R})$ is the steady-state density matrix at each nuclear position $\mathbf{R}$, i.e. $\partial_{t}\rhoss(\mathbf{R})=-i[\hat{H}(\mathbf{R}),\rhoss(\mathbf{R})]=0$. In Eq. (\ref{eq:general_ft}), the partial derivatives with respect to the nuclear coordinates $\mu$, $\nu$ (i.e. $\partial_{\mu}$, $\partial_{\nu}$) operate only on the operator directly to the right, and this convention will be used throughout the letter below. The trace is taken over all the electronic degrees of freedom.

Equation (\ref{eq:general_ft}) can be further simplified when a general non-interacting Hamiltonian, $\hat{H}=\sum_{pq}\mathcal{H}_{pq}\hat{d}_{p}^{\dagger}\hat{d}_{q}+U(\mathbf{R})$ ($\hat{d}_{p}^{\dagger}$/$\hat{d}_{p}$ creates/annihilates an electron in orbital $p$, and $U(\mathbf{R})$ is a pure nuclear potential energy), is considered. The result is:
\begin{align}
\gamma_{\mu\nu}=-\frac{1}{2\pi}\int_{-\infty}^{\infty}d\epsilon\,\Tr{\partial_{\mu}\mathcal{H}\Ropt{\mathcal{G}}\partial_{\nu}\sigma_{\mathrm{ss}}\Aopt{\mathcal{G}}},\label{eq:ft_noninteracting}
\end{align}
where $\RAopt{\mathcal{G}}=(\epsilon-\mathcal{H}\pm i\eta)^{-1}$ are retarded/advanced Green's functions of the electrons, and $(\sigma_{\mathrm{ss}})_{pq}=\Tr{\rhoss\hat{d}_{p}^{\dagger}\hat{d}_{q}}$ is the steady-state density matrix. In general, $\sigma_{\mathrm{ss}}$ is a complicated function of $\hat{H}$ so that the analytic derivative of the steady-state distribution (e.g. a Boltzmann distribution which depends on $\mathbf{R}$) is hard to calculate.  Nevertheless, from this expression, one can show that $\gamma_{\mu\nu}^{\mathrm{A}}$ vanishes when $\hat{H}$ is real-valued and the system is in equilibrium; see SM \ref{si:no_gamma_A_for_real_H_equil} for a proof. In practice, we will use a nonequilibrium Green's function technique to treat $\sigma_{\mathrm{ss}}$. Notice that $\sigma_{\mathrm{ss}}$ can be expressed in terms of the lesser Green's function in the frequency domain,
\begin{align*}
(\sigma_{\mathrm{ss}})_{pq}=-i\mathcal{G}_{pq}^{<}(t,t)=-i\int\frac{d\epsilon'}{2\pi}\mathcal{G}_{pq}^{<}(\epsilon'),
\end{align*}
where the lesser Green's function $\mathcal{G}^{<}(t_{1},t_{2})=i\Tr{\rhoss\hat{d}_{p}^{\dagger}(t_{2})\hat{d}_{q}(t_{1})}$ is defined as usual. Since $\sigma_{\mathrm{ss}}$ applies at steady state, $\mathcal{G}^{<}(\epsilon)$ can be calculated by the Keldysh equation\cite{haug2008quantum},
\begin{align*}
\mathcal{G}^{<}=\Ropt{\mathcal{G}}\Pi^{<}\Aopt{\mathcal{G}},
\end{align*}
where $\Pi^{<}$ is the lesser self energy of the electrons. Here we artificially imagine very weakly embedding the total electronic Hamiltonian $\hat{H}$ in some fictitious surrounding so that $\Pi^{<}$ can be evaluated. Because, the interaction between $\hat{H}$ and the surrounding is taken as infinitesimally small, all properties we derive will not be affected by the introduction of an artificial surrounding. In general, $\mathcal{G}^{\mathrm{R},\mathrm{A}}$ and $\Pi^{<}$ can all possibly depend on $\mathbf{R}$. However, we further assume that $\Pi^{<}$ is independent of $\mathbf{R}$ so that $\gamma_{\mu\nu}$ can be substantially simplified. This assumption can be realized by imaging that the total electronic Hamiltonian $\hat{H}$ is composed of two parts: one that depends on $\mathbf{R}$ and the other that does not depend on $\mathbf{R}$, and only the later weakly couples to the surrounding. According to Ref. \citenum{dou2018universality}, using the assumptions above, one can transform Eq. (\ref{eq:ft_noninteracting}) into the following expression,
\begin{align}
\gamma_{\mu\nu}=\int_{-\infty}^{\infty}\frac{d\epsilon}{2\pi}\Tr{\partial_{\mu}\mathcal{H}\partial_{\epsilon}\Ropt{\mathcal{G}}\partial_{\nu}\mathcal{H}\mathcal{G}^{<}}+\mathrm{H.c.},\label{eq:ft_noninteracting_nc}
\end{align}
Note that the potential $U(\mathbf{R})$ does not contribute to the friction tensor.

Finally, if we make the Condon approximation and assume that the coupling between the system and the bath $V_{m,k\alpha}$ is independent of the nuclear coordinate $\mathbf{R}$, the friction tensor in Eq. (\ref{eq:ft_noninteracting_nc}) becomes Eq. (\ref{eq:ft_non_condon}) in the main text\cite{dou2018universality}.

\section{Positive Definiteness of the Electronic Friction Tensor}\label{si:positive_definite_ft_equil}
In this section, we prove that both the electronic friction tensor $\gamma_{\mu\nu}$ and its symmetric part $\gamma_{\mu\nu}^{\mathrm{S}}$ are positive definite when the system is in equilibrium. That is, $\sum_{\mu\nu}X_{\mu}\gamma_{\mu\nu}X_{\nu}>0$ for arbitrary real vectors $\mathbf{X}\neq0$. We start from the identity (which can be easily proved by the fundamental theorem of calculus),
\begin{align*}
e^{-t\hat{A}}\frac{d}{d\lambda}e^{t\hat{A}}=\int_{0}^{t}ds\,e^{-s\hat{A}}\frac{d\hat{A}}{d\lambda}e^{s\hat{A}},
\end{align*}
where $\hat{A}$ is an arbitrary operator and $t$ is real. By replacing $\hat{A}=\ln{\rho}$ and $t=1$, we obtain the following expression for the derivative of the steady-state density matrix (for notational simplicity, we discard the subscript $\mathrm{ss}$ of $\rhoss$ and the hat symbol $\hat{}$ for operators in this section),
\begin{align}
\frac{d}{d\lambda}\rho=\int_{0}^{1}ds\,\rho^{1-s}\frac{d\ln{\rho}}{d\lambda}\rho^{s}.\label{eq:kubo_transf}
\end{align}
Therefore, Eq. (\ref{eq:general_ft}) can be recast into
\begin{align*}
\gamma_{\mu\nu}=-\int_{0}^{\infty}dt\int_{0}^{1}ds\,\Tr{e^{-iHt}\rho^{1-s}\partial_{\nu}(\ln{\rho})\rho^{s}e^{iHt}\partial_{\mu}H}.
\end{align*}
Next we apply the equilibrium condition for the steady-state density matrix, $\rho=e^{-\beta H}/Z$ where $Z\equiv\Tr{e^{-\beta H}}$ is the partition function. This condition is equivalent to $\partial_{\mu}H=\left(-\partial_{\mu}\ln{\rho}-\partial_{\mu}Z/Z\right)/\beta$. Accordingly,
\begin{align}
\gamma_{\mu\nu}=\frac{1}{\beta}\int_{0}^{\infty}dt\int_{0}^{1}ds\,\Tr{e^{-iHt}\rho^{1-s}\partial_{\nu}(\ln{\rho})\rho^{s}e^{iHt}\partial_{\mu}(\ln{\rho})}.\label{eq:ft_for_proving_positive_definiteness}
\end{align}
Notice that another term proportional to $\partial_{\mu}Z/Z$ vanishes since the integrand becomes $\Tr{\partial_{\nu}\rho}=\partial_{\nu}\Tr{\rho}=0$ after using Eq. (\ref{eq:kubo_transf}).

Now, in order to see the structure of $\gamma_{\mu\nu}$ more easily, we rewrite the trace in the Lehmann representation, $H\vert a\rangle=E_{a}\vert a\rangle$,
\begin{align*}
\gamma_{\mu\nu}=&\sum_{ab}\frac{1}{\beta}\int_{0}^{\infty}dt\int_{0}^{1}ds\,e^{-iE_{a}t}
\left(\frac{e^{-\beta E_{a}}}{Z}\right)^{1-s}\langle a\vert\partial_{\nu}(\ln{\rho})\vert b\rangle\left(\frac{e^{-\beta E_{b}}}{Z}\right)^{s}e^{iE_{b}t}\langle b\vert\partial_{\mu}(\ln{\rho})\vert a\rangle\\
=&\frac{1}{\beta}\sum_{ab}\int_{0}^{1}ds\,\rho_{a}^{1-s}\rho_{b}^{s}\langle a\vert\partial_{\nu}(\ln{\rho})\vert b\rangle\langle b\vert\partial_{\mu}(\ln{\rho})\vert a\rangle\frac{i}{E_{b}-E_{a}+i\eta},
\end{align*}
where $\rho_{a}\equiv e^{-\beta E_{a}}/Z$ (same for $\rho_{b}$), and $\eta\rightarrow0^{+}$. We further split $\gamma_{\mu\nu}$ into the symmetric part and the antisymmetric part, and analyze the symmetric part first,
\begin{align}
\gamma_{\mu\nu}^{\mathrm{S}}=&\frac{1}{2}\sum_{ab}\frac{1}{\beta}\int_{0}^{1}ds\,\rho_{a}^{1-s}\rho_{b}^{s}\langle a\vert\partial_{\nu}(\ln{\rho})\vert b\rangle\langle b\vert\partial_{\mu}(\ln{\rho})\vert a\rangle i\left(\frac{1}{E_{b}-E_{a}+i\eta}-\frac{1}{E_{b}-E_{a}-i\eta}\right)\notag\\
=&\pi\sum_{ab}\frac{1}{\beta}\int_{0}^{1}ds\,\rho_{a}^{1-s}\rho_{b}^{s}\langle a\vert\partial_{\nu}(\ln{\rho})\vert b\rangle\langle b\vert\partial_{\mu}(\ln{\rho})\vert a\rangle\delta(E_{b}-E_{a})\notag\\
=&\frac{\pi}{\beta}\sum_{ab}\rho_{a}\langle a\vert\partial_{\nu}(\ln{\rho})\vert b\rangle\langle b\vert\partial_{\mu}(\ln{\rho})\vert a\rangle\delta(E_{a}-E_{b}).\label{eq:gamma_s}
\end{align}
Thus,
\begin{align*}
\sum_{\mu\nu}X_{\mu}\gamma_{\mu\nu}^{\mathrm{S}}X_{\nu}=\frac{\pi}{\beta}\sum_{ab}\rho_{a}\abs{\langle a\vert\sum_{\nu}X_{\nu}\partial_{\nu}(\ln{\rho})\vert b\rangle}^{2}\delta(E_{b}-E_{a})>0.
\end{align*}
Second, since the antisymmetric part $\gamma_{\mu\nu}^{\mathrm{A}}$ is also real,
\begin{align*}
\sum_{\mu\nu}X_{\mu}\gamma_{\mu\nu}X_{\nu}
=\sum_{\mu\nu}X_{\mu}\left(\gamma_{\mu\nu}^{\mathrm{S}}+\gamma_{\mu\nu}^{\mathrm{A}}\right)X_{\nu}
=\sum_{\mu\nu}X_{\mu}\gamma_{\mu\nu}^{\mathrm{S}}X_{\nu}>0.
\end{align*}
Hence, $\gamma_{\mu\nu}$ is positive definite at equilibrium.

\section{Fluctuation-Dissipation Theorem Based on the QCLE}\label{si:fd_theorem_based_on_qcle}
In this section, we further investigate the fluctuation-dissipation theorem based on the QCLE. Note that the derivation published in the SM of Ref. \citenum{dou2017born} erroneously divides by zero, but the final result is correct for a real-valued Hamiltonian (as we will now show). 

In Ref.  \citenum{dou2017born}, a Fokker-Planck equation (or more specifically a Kramer's equation) based on an analysis of the QCLE in the adiabatic theorem was derived along with the corresponding random force $\delta\hat{F}_{\mu}$ and correlation function $\bar{D}_{\mu\nu}^{\mathrm{S}}$.  
\begin{align*}
\delta\hat{F}_{\mu}=&-\partial_{\mu}\hat{H}+\Tr{\partial_{\mu}\hat{H}\rhoss},\\
\bar{D}_{\mu\nu}^{\mathrm{S}}=&\frac{1}{2}\int_{0}^{\infty}dt\,\Tr{e^{i\hat{H}t}\delta\hat{F}_{\mu}e^{-i\hat{H}t}\left(\delta\hat{F}_{\nu}\rhoss+\rhoss\delta\hat{F}_{\nu}\right)}.
\end{align*}
(We follow the same notation as in Ref. \citenum{dou2017born}, even though $\bar{D}_{\mu\nu}^{\mathrm{S}}$ is not symmetric when a complex-valued Hamiltonian is considered here.)
Since $\partial_{\nu}\ln{\rhoss}=\beta\delta\hat{F}_{\nu}$, we can rewrite Eq. (\ref{eq:ft_for_proving_positive_definiteness}),
\begin{align*}
\gamma_{\mu\nu}=\beta\int_{0}^{\infty}dt\int_{0}^{1}ds\Tr{\left(\rhoss\right)^{1-s}\delta\hat{F}_{\nu}\left(\rhoss\right)^{s}\delta\hat{F}_{\mu}(t)},
\end{align*}
where $\delta\hat{F}_{\mu}(t)$ is written in Heisenberg picture. We then express both $\gamma_{\mu\nu}$ and $\bar{D}_{\mu\nu}^{\mathrm{S}}$ in the Lehmann representation, obtaining
\begin{align}
\gamma_{\mu\nu}=&\beta\sum_{ab}\int_{0}^{\infty}dt\int_{0}^{1}ds\,\rho_{a}^{1-s}\rho_{b}^{s}\langle a\vert\delta\hat{F}_{\nu}\vert b\rangle\langle b\vert\delta\hat{F}_{\mu}(t)\vert a\rangle\\
=&\beta\sum_{ab}\frac{\rho_{b}-\rho_{a}}{\beta(E_{a}-E_{b})}\int_{0}^{\infty}dt\,\langle a\vert\delta\hat{F}_{\nu}\vert b\rangle\langle b\vert\delta\hat{F}_{\mu}(t)\vert a\rangle,\label{eq:fd_theorem_gamma}\\
\bar{D}_{\mu\nu}^{\mathrm{S}}=&\sum_{ab}\frac{\rho_{a}+\rho_{b}}{2}\int_{0}^{\infty}dt\,\langle a\vert\delta\hat{F}_{\nu}\vert b\rangle\langle b\vert\delta\hat{F}_{\mu}(t)\vert a\rangle.\label{eq:fd_theorem_D}
\end{align}
At this point, recall that, according to a Kramer's equation, a particle's equation of motion does not depend on the antisymmetric component of the random force correlation function. In other words, if $\mathcal{A}$ is the phase space density of a particle near a surface, the equation of motion for $\mathcal{A}$ satisfies\cite{dou2017born}:
\begin{align}
\partial_{t}\mathcal{A}=-\sum_{\alpha}\frac{P_{\alpha}}{m_{\alpha}}\partial_{\alpha}\mathcal{A}-\sum_{\alpha}F_{\alpha}\frac{\partial\mathcal{A}}{\partial P_{\alpha}}+\sum_{\alpha\nu}\gamma_{\alpha\nu}\frac{\partial}{\partial P_{\alpha}}\left(\frac{P_{\nu}}{m_{\nu}}\mathcal{A}\right)+\sum_{\alpha\nu}\bar{D}_{\alpha\nu}^{\mathrm{S}}\frac{\partial^{2}\mathcal{A}}{\partial P_{\alpha}\partial P_{\nu}}.
\end{align}
Thus, the physical meaning of the antisymmetric component of  $\bar{D}_{\mu\nu}^{\mathrm{S}}$ is not clear. Perhaps not surprisingly, then, 
Equations (\ref{eq:fd_theorem_gamma}) and (\ref{eq:fd_theorem_D}) do not satisfy
a ``fluctuation-dissipation theorem'', $\gamma_{\mu\nu} \ne \beta\bar{D}_{\mu\nu}^{\mathrm{S}}$.  However, a 
valid fluctuation-dissipation theorem condition can be established 
at equilibrium if we consider only the 
symmetric component of the random force and friction. To do so, we further integrate out the time variable in $\bar{D}_{\mu\nu}^{\mathrm{S}}$,
\begin{align*}
\bar{D}_{\mu\nu}^{\mathrm{S}}=\sum_{ab}\frac{\rho_{a}+\rho_{b}}{2}\frac{i}{E_{b}-E_{a}+i\eta}\langle a\vert\delta\hat{F}_{\nu}\vert b\rangle\langle b\vert\delta\hat{F}_{\mu}\vert a\rangle,
\end{align*}
and then ``symmetrize'' $\bar{D}_{\mu\nu}^{\mathrm{S}}$:
\begin{align*}
&\frac{1}{2}\left(\bar{D}_{\mu\nu}^{\mathrm{S}}+\bar{D}_{\nu\mu}^{\mathrm{S}}\right)\\
=&\frac{1}{2}\sum_{ab}\frac{\rho_{a}+\rho_{b}}{2}\left(\frac{i}{E_{b}-E_{a}+i\eta}+\frac{i}{E_{a}-E_{b}+i\eta}\right)\langle a\vert\delta\hat{F}_{\nu}\vert b\rangle\langle b\vert\delta\hat{F}_{\mu}\vert a\rangle\\
=&\pi\sum_{ab}\rho_{a}\langle a\vert\delta\hat{F}_{\nu}\vert b\rangle\langle b\vert\delta\hat{F}_{\mu}\vert a\rangle\delta(E_{a}-E_{b}),
\end{align*}
which is equal to $\gamma_{\mu\nu}^{\mathrm{S}}/\beta$ when the system is in equilibrium (please compare to Eq. (\ref{eq:gamma_s})). 

In SM \ref{si:no_gamma_A_for_real_H_equil}, we will show that when a real-valued Hamiltonian is considered and the system is in equilibrium, the antisymmetric friction tensor $\gamma_{\mu\nu}^{\mathrm{A}}$ vanishes. In this situation, $\gamma_{\mu\nu}=\gamma_{\mu\nu}^{\mathrm{S}}=\beta\bar{D}_{\mu\nu}^{\mathrm{S}}$.

\section{No Antisymmetric Friction Tensor $\gamma_{\mu\nu}^{\mathrm{A}}$ When the Hamiltonian Is Real and the System Is in Equilibrium}\label{si:no_gamma_A_for_real_H_equil}
In this section, we show that the antisymmetric friction $\gamma_{\mu\nu}^{\mathrm{A}}$ vanishes when the Hamiltonian is real and the system is in equilibrium (for a non-interacting Hamiltonian). We start from Eq. (\ref{eq:ft_noninteracting}),
\begin{align}
\gamma_{\alpha\nu}=-\sum_{kl}\int_{-\infty}^{\infty}\frac{d\epsilon}{2\pi}\langle k\vert\partial_{\alpha}\mathcal{H}\vert l\rangle\frac{1}{\epsilon-\epsilon_{l}-i\eta}\langle l\vert\partial_{\nu}\sigma^{\mathrm{ss}}\vert k\rangle\frac{1}{\epsilon-\epsilon_{k}+i\eta},\label{eq:ft_noninteracting_expand}
\end{align}
where $\eta\rightarrow0^{+}$, and
\begin{align*}
\mathcal{H}=\sum_{k}\epsilon_{k}\vert k\rangle\langle k\vert.
\end{align*}
At equilibrium we have
\begin{align*}
\sigma^{\mathrm{ss}}=\sum_{k}P(\epsilon_{k})\vert k\rangle\langle k\vert,
\end{align*}
where $P(\epsilon_{k})=e^{-\beta\epsilon_{k}}/\sum_{k}e^{-\beta\epsilon_{k}}$ is the Boltzmann distribution. We focus on $\gamma_{\alpha\nu}^{\mathrm{A}}\propto\gamma_{\alpha\nu}-\gamma_{\nu\alpha}$. We divide the summation $\sum_{kl}$ in Eq. (\ref{eq:ft_noninteracting_expand}) into three cases:
\begin{align}
\sum_{kl}=\sum_{k=l}+\sum_{k\neq l,\epsilon_{k}\neq\epsilon_{l}}+\sum_{k\neq l,\epsilon_{k}=\epsilon_{l}}.\label{eq:sums}
\end{align}
Also, we utilize the following two identities to replace $\langle l\vert\partial_{\nu}\sigma^{\mathrm{ss}}\vert k\rangle$ in Eq. (\ref{eq:ft_noninteracting_expand}):
\begin{align}
&\partial_{\nu}\epsilon_{k}\delta_{lk}=(\epsilon_{l}-\epsilon_{k})\langle l\vert\partial_{\nu}\vert k\rangle+\langle l\vert\partial_{\nu}h\vert k\rangle,\\
&\langle l\vert\partial_{\nu}\sigma^{\mathrm{ss}}\vert k\rangle=\partial_{\nu}\epsilon_{k}\frac{\partial f(\epsilon_{k})}{\partial\epsilon_{k}}\delta_{kl}+\langle l\vert\partial_{\nu}\vert k\rangle\left(f(\epsilon_{k})-f(\epsilon_{l})\right).
\end{align}
As a result,
\begin{align}
\gamma_{\alpha\nu}^{\mathrm{A}}\propto
&\sum_{k\neq l,\epsilon_{k}\neq\epsilon_{l}}\int\frac{d\epsilon}{2\pi}\langle k\vert\partial_{\alpha}\mathcal{H}\vert l\rangle\frac{1}{\epsilon-\epsilon_{l}-i\eta}\left\{-\langle l\vert\partial_{\nu}\mathcal{H}\vert k\rangle\frac{f(\epsilon_{k})-f(\epsilon_{l})}{\epsilon_{l}-\epsilon_{k}}\right\}\frac{1}{\epsilon-\epsilon_{k}+i\eta}\notag\\
&+\sum_{k\neq l,\epsilon_{k}=\epsilon_{l}}\int\frac{d\epsilon}{2\pi}\left\{0\right\}\frac{1}{\epsilon-\epsilon_{l}-i\eta}\left\{0\right\}\frac{1}{\epsilon-\epsilon_{k}+i\eta}\notag\\
&-(\alpha\leftrightarrow\nu)\label{eq:three_sums}.
\end{align}
Note that the diagonal term (the first summation $\sum_{k=l}$ in Eq. (\ref{eq:sums})) does not contribute to $\gamma_{\alpha\nu}^{\mathrm{A}}$. Similarly, the second line of Eq. (\ref{eq:three_sums}) is zero, which is consistent with the assumption we made in deriving the friction tensor Eq. (\ref{eq:general_ft}): in order to achieve a well-defined unique steady state $\hat{\rho}_{\mathrm{ss}}$ we presume there are no degenerate states.

Next, let's focus on the only contributing summation in Eq. (\ref{eq:three_sums}). By using the identities,
\begin{align*}
&\frac{c}{2\pi}\int_{-\infty}^{\infty}dy\,e^{icxy}=\delta(x),\\
&\theta(t_{1}-t_{2})=i\int_{-\infty}^{\infty}\frac{d\omega}{2\pi}\frac{e^{-i\omega(t_{1}-t_{2})}}{\omega+i\eta},
\end{align*}
we can derive the following expression:
\begin{align}
\int_{0}^{\infty}dt\,e^{i(E_{b}-E_{a})t}=&\int_{-\infty}^{\infty}dt\,\theta(t)e^{i(E_{b}-E_{a})t}=\int_{-\infty}^{\infty}dt\,\left(i\int_{-\infty}^{\infty}\frac{d\omega}{2\pi}\frac{e^{-i\omega t}}{\omega+i\eta}\right)e^{i(E_{a}-E_{b})t}\notag\\
=&i\int_{-\infty}^{\infty}\frac{d\omega}{2\pi}\frac{1}{\omega+i\eta}\int_{-\infty}^{\infty}dt\,e^{i(E_{b}-E_{a}-\omega)t}=\frac{i}{E_{b}-E_{a}+i\eta}.\label{eq:identity3}
\end{align}
We then use Eq. (\ref{eq:identity3}) to rewrite the first line in Eq. (\ref{eq:three_sums}), obtaining
\begin{align}
&\sum_{k\neq l,\epsilon_{k}\neq\epsilon_{l}}\int\frac{d\epsilon}{2\pi}\langle k\vert\partial_{\alpha}\mathcal{H}\vert l\rangle i\int_{0}^{\infty}dt\,e^{-i(\epsilon-\epsilon_{l})t}\left\{-\langle l\vert\partial_{\nu}\mathcal{H}\vert k\rangle\frac{f(\epsilon_{k})-f(\epsilon_{l})}{\epsilon_{l}-\epsilon_{k}}\right\}(-i)\int_{0}^{\infty}dt'\,e^{i(\epsilon-\epsilon_{k})t'}\notag\\
=&\sum_{k\neq l,\epsilon_{k}\neq\epsilon_{l}}\int\frac{d\epsilon}{2\pi}\int_{0}^{\infty}dt\int_{0}^{\infty}dt'\,e^{i\epsilon(t'-t)}e^{i\epsilon_{l}t}e^{-i\epsilon_{k}t'}\langle k\vert\partial_{\alpha}\mathcal{H}\vert l\rangle\left\{-\langle l\vert\partial_{\nu}\mathcal{H}\vert k\rangle\frac{f(\epsilon_{k})-f(\epsilon_{l})}{\epsilon_{l}-\epsilon_{k}}\right\}\notag\\
=&\sum_{k\neq l,\epsilon_{k}\neq\epsilon_{l}}\int_{0}^{\infty}dt\int_{0}^{\infty}dt'\,\delta(t'-t)e^{i\epsilon_{l}t}e^{-i\epsilon_{k}t'}\langle k\vert\partial_{\alpha}\mathcal{H}\vert l\rangle\left\{-\langle l\vert\partial_{\nu}\mathcal{H}\vert k\rangle\frac{f(\epsilon_{k})-f(\epsilon_{l})}{\epsilon_{l}-\epsilon_{k}}\right\}\notag\\
=&\sum_{k\neq l,\epsilon_{k}\neq\epsilon_{l}}\int_{0}^{\infty}dt\,e^{i(\epsilon_{l}-\epsilon_{k})t}\langle k\vert\partial_{\alpha}\mathcal{H}\vert l\rangle\left\{-\langle l\vert\partial_{\nu}\mathcal{H}\vert k\rangle\frac{f(\epsilon_{k})-f(\epsilon_{l})}{\epsilon_{l}-\epsilon_{k}}\right\}\notag\\
=&\sum_{k\neq l,\epsilon_{k}\neq\epsilon_{l}}\frac{i}{\epsilon_{l}-\epsilon_{k}+i\eta}\langle k\vert\partial_{\alpha}\mathcal{H}\vert l\rangle\left\{-\langle l\vert\partial_{\nu}\mathcal{H}\vert k\rangle\frac{f(\epsilon_{k})-f(\epsilon_{l})}{\epsilon_{l}-\epsilon_{k}}\right\}.\label{eq:three_sum_2nd_recast}
\end{align}
By using the identity,
\begin{align*}
\frac{1}{\omega\pm i\eta}=\mathcal{P}\frac{1}{\omega}\mp i\pi\delta(\omega),
\end{align*}
we can recast Eq. (\ref{eq:three_sum_2nd_recast}) to get
\begin{align}
&i\sum_{k\neq l,\epsilon_{k}\neq\epsilon_{l}}\mathcal{P}\frac{1}{\epsilon_{l}-\epsilon_{k}}\langle k\vert\partial_{\alpha}\mathcal{H}\vert l\rangle\left\{-\langle l\vert\partial_{\nu}\mathcal{H}\vert k\rangle\frac{f(\epsilon_{k})-f(\epsilon_{l})}{\epsilon_{l}-\epsilon_{k}}\right\}\notag\\
&+\pi\sum_{k\neq l,\epsilon_{k}\neq\epsilon_{l}}\delta(\epsilon_{l}-\epsilon_{k})\langle k\vert\partial_{\alpha}\mathcal{H}\vert l\rangle\left\{-\langle l\vert\partial_{\nu}\mathcal{H}\vert k\rangle\frac{f(\epsilon_{k})-f(\epsilon_{l})}{\epsilon_{l}-\epsilon_{k}}\right\}.
\end{align}
Apparently, the second term is $0$ since $\epsilon_{k}\neq\epsilon_{l}$. Also, the first term will not contribute if the Hamiltonian is real, since the friction tensor $\gamma_{\mu\nu}$ must be real and so we only need the real part of $\gamma_{\mu\nu}^{\mathrm{A}}$. Therefore, we have proven that the antisymmetric friction tensor vanishes when a real Hamiltonian is considered and the system is in equilibrium.

As a side note, Eq. (\ref{eq:three_sums}) can be recast in the following form,
\begin{align*}
\gamma_{\alpha\nu}^{\mathrm{A}}\propto-\sum_{k\neq l,\epsilon_{k}\neq\epsilon_{l}}2\mathfrak{Im}\left\{d_{kl}^{\alpha}d_{lk}^{\nu}\right\}\left[f(\epsilon_{k})-f(\epsilon_{l})\right],
\end{align*}
where $d_{kl}^{\alpha}\equiv\langle k\vert\partial_{\alpha}\vert l\rangle$ is the derivative coupling. As a result, the Lorentz force $-\sum_{\nu}\gamma_{\alpha\nu}^{\mathrm{A}}\dot{R}_{\nu}$ (which is defined in the main body of the text) becomes the normal Berry force weighted by the Fermi distributions:
\begin{align*}
-\sum_{\nu}\dot{R}_{\nu}\gamma_{\alpha\nu}^{\mathrm{A}}=
2\mathfrak{Im}\left\{\sum_{k\neq l,\epsilon_{k}\neq\epsilon_{l}}\left(\dot{\mathbf{R}}\cdot\mathbf{d}_{lk}\right)d_{kl}^{\alpha}\right\}
\left[f(\epsilon_{k})-f(\epsilon_{l})\right].
\end{align*}

\section{Schematic Picture of the Molecular Junction Hamiltonian (Equilibrium Case)}\label{si:schematic_pic}
\begin{figure}[!h]
\centering
\includegraphics[width=.55\textwidth]{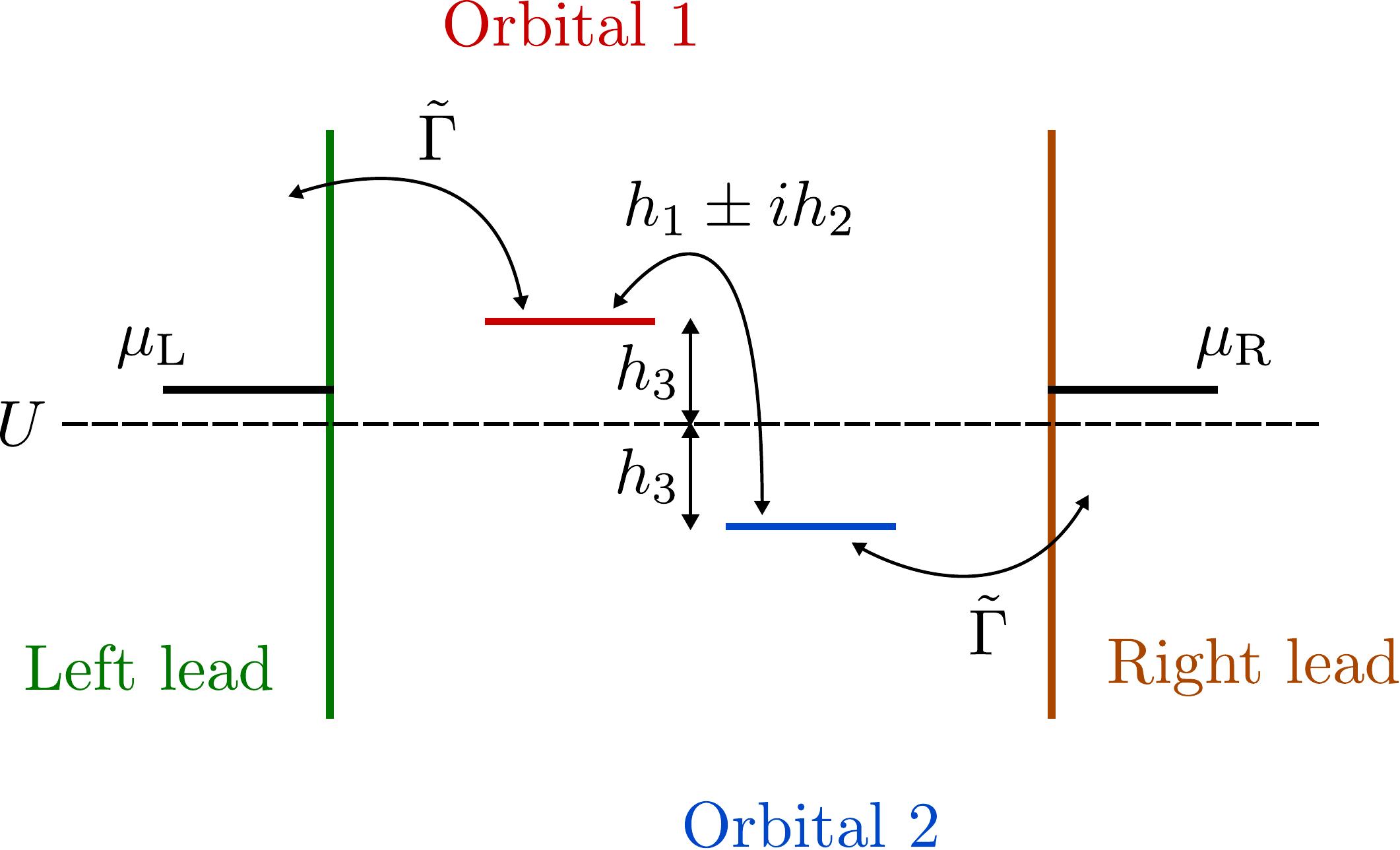}
\caption{A schematic picture of the molecular junction Hamiltonian used in the main body of the text. Here we plot the equilibrium case where $\mu_{\mathrm{L}}=\mu_{\mathrm{R}}$.}
\end{figure}

\section{Deriving Equations (\ref{eq:ft_tls})-(\ref{eq:ft_tls_antisym})}\label{si:ft_calculation_details}
Here we provide a few details as far as deriving Eqs. (\ref{eq:ft_tls})-(\ref{eq:ft_tls_antisym}). Under the approximations made in the main text, the retarded Green's function is
\begin{align*}
\Ropt{G}=\frac{1}{\epsilon-h^{\mathrm{s}}-\Ropt{\Sigma}}=\frac{1}{\tilde{\epsilon}-\mathbf{h}\cdot\bm{\sigma}}=\frac{1}{\tilde{\epsilon}^{2}-h^{2}}\left(\tilde{\epsilon}+\mathbf{h}\cdot\bm{\sigma}\right),
\end{align*}
As a result, the lesser Green's function in Eq. (\ref{eq:ft_non_condon}) can be calculated,
\begin{align*}
G^{<}=\Ropt{G}\Sigma^{<}\Aopt{G}=i\tilde{\Gamma}\abs{\frac{1}{\tilde{\epsilon}^{2}-h^{2}}}^{2}\left(\kappa_{0}+\bm{\kappa}\cdot\bm{\sigma}\right),
\end{align*}
where $\Aopt{G}=\left(\Ropt{G}\right)^{\dagger}$.

We then calculate the overall trace in Eq. (\ref{eq:ft_non_condon}) (recall that $\partial_{\epsilon}\Ropt{G}=-\Ropt{G}\Ropt{G}$),
\begin{align*}
&C\,\Tr{\partial_{\mu}\mathbf{h}\cdot\bm{\sigma}\left(\tilde{\epsilon}^{2}+h^{2}+2\tilde{\epsilon}\mathbf{h}\cdot\bm{\sigma}\right)\partial_{\nu}\mathbf{h}\cdot\bm{\sigma}\left(\kappa_{0}+\bm{\kappa}\cdot\bm{\sigma}\right)}\\
=&2C\big\{2\tilde{\epsilon}\left(\partial_{\mu}\mathbf{h}\cdot\mathbf{h}\right)\left(\partial_{\nu}\mathbf{h}\cdot\bm{\kappa}\right)+\left(\tilde{\epsilon}^{2}+h^{2}\right)\kappa_{0}\partial_{\mu}\mathbf{h}\cdot\partial_{\nu}\mathbf{h}\\
&+i\left(\tilde{\epsilon}^{2}+h^{2}\right)\partial_{\mu}\mathbf{h}\cdot\left(\partial_{\nu}\mathbf{h}\times\bm{\kappa}\right)+i2\tilde{\epsilon}\kappa_{0}\left(\partial_{\mu}\mathbf{h}\times\mathbf{h}\right)\cdot\partial_{\nu}\mathbf{h}\\
&-2\tilde{\epsilon}\left(\partial_{\mu}\mathbf{h}\times\mathbf{h}\right)\cdot\left(\partial_{\nu}\mathbf{h}\times\bm{\kappa}\right)\big\},
\end{align*}
In order to calculate the friction tensor, we only need to consider the real part of the trace above, resulting in the final expression of the friction tensor in Eqs. (\ref{eq:ft_tls})-(\ref{eq:ft_tls_antisym}).

\section{$h^{\mathrm{s}}$ with a Conical Intersection (Equilibrium)}\label{si:ci}
In this section, we prove that the antisymmetric friction tensor $\gamma_{xy}^{A}$ vanishes when a typical conical intersection (with only linear dependence on the nuclear coordinates) is considered. We then model a simple $E\otimes\epsilon$ Jahn-Teller system in the presence of spin-orbit coupling and a nuclear bath to demonstrate that no significant enhancement of the relative strength of the antisymmetric friction tensor arises from the presence of a conical intersection (which does provide infinite derivative couplings for an isolated system). To prove these points, assume that a conical intersection is located at $(0,0)$. The most general second order Hamiltonian is
\begin{align*}
\mathbf{h}^{\mathrm{s}} &= \mathbf{h}(x,y) \cdot \bm{\sigma}, \\
\mathbf{h}(x,y) &= \mathbf{P}x^{2}+\mathbf{Q}xy+\mathbf{R}y^{2}+\mathbf{S}x+\mathbf{T}y,
\end{align*}
where $\mathbf{P}$, $\mathbf{Q}$, $\mathbf{R}$, $\mathbf{S}$ and $\mathbf{T}$ are constant vectors.
If only two of these constant vectors are nonzero, the antisymmetric friction tensor $\gamma_{\mu\nu}^{\mathrm{A}}$ (which depends on $\mathbf{h}\cdot\left(\partial_{\mu}\mathbf{h}\times\partial_{\nu}\mathbf{h}\right)$) must vanish. This fact indicates that, for a typical conical intersection with a consideration of only linear vibronic terms (only $\mathbf{S}$ and $\mathbf{T}$ are nonzero), there is no antisymmetric friction tensor. Moreover, when there are no linear terms (only $\mathbf{P}$, $\mathbf{Q}$ and $\mathbf{R}$ are nonzero as in a Renner-Teller intersection\cite{grosso2014solid}), $\gamma_{\mu\nu}^{\mathrm{A}}$ must still disappear.

Finally, in order to derive a nonzero $\gamma_{\mu\nu}^{\mathrm{A}}$, we must include a linear complex coupling on top of a second order real-valued $E\otimes\epsilon$ Jahn--Teller system,
\begin{align*}
\mathbf{h}^{1}=
\begin{pmatrix}
y^{2}-x^{2} & 2xy\\
2xy & x^{2}-y^{2}
\end{pmatrix}
+A
\begin{pmatrix}
-y & x\\
x & y
\end{pmatrix}
+B
\begin{pmatrix}
0 & -iy\\
iy & 0
\end{pmatrix},
\end{align*}
This Hamiltonian can be experimentally realized as a regular triangular molecule with two degenerate electronic states interacting with a doublet of vibrational states (up to quadratic order) with a spin-orbit coupling between the $p_{x}$ and $p_{y}$ orbitals included.

\begin{figure}[!h]
\centering
\includegraphics[width=.55\textwidth]{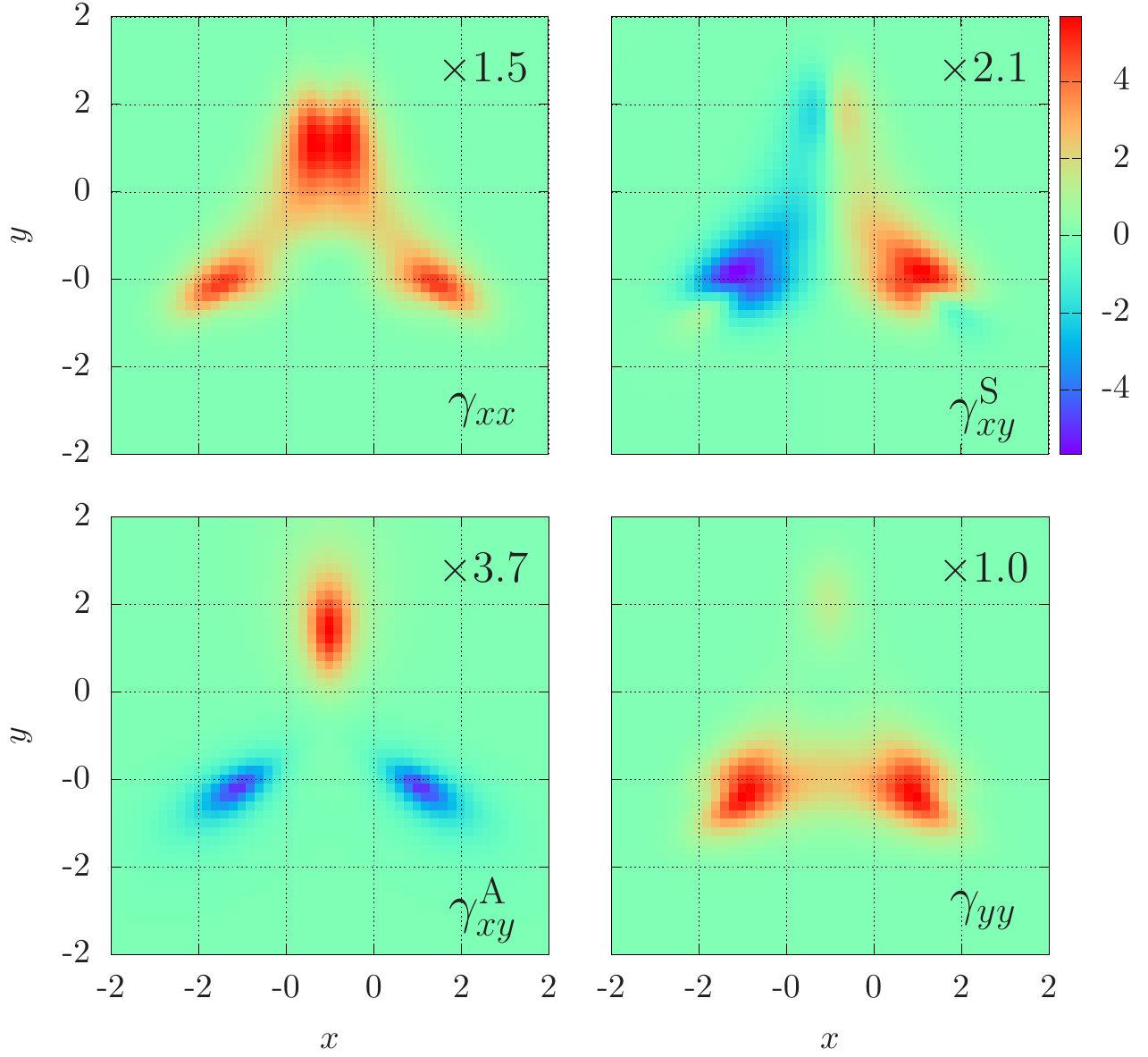}
\caption{Friction tensor calculation results near a conical intersection: $\gamma_{xx}$ (top left), $\gamma_{xy}^{\mathrm{S}}$ (top right), $\gamma_{xy}^{\mathrm{A}}$ (bottom left) and $\gamma_{yy}$ (bottom right). Parameters: $\tilde{\Gamma}=1$, $\mu_{\mathrm{R}}=\mu_{\mathrm{L}}=0$, $\beta=2$, $A=1$, $B=0.5$. Note that no enhancement of the relative strength of the antisymmetric pseudomagnetic field is caused by a conical intersection. Also, near the conical intersection where the second order vibronic terms can be ignored, $\gamma_{xy}^{\mathrm{A}}$ vanishes.\label{fig:jahn_teller_quadratic}}
\end{figure}

Figure \ref{fig:jahn_teller_quadratic} plots the corresponding friction tensor results, and the ratio $\abs{\gamma_{yy}^{\mathrm{S}}/\gamma}$. After scanning a reasonable set of parameters, we have never, in practice, been able to find a Hamiltonian where this ratio is more than $~25\%$ in the vicinity of the origin. Thus, we do not observe any enhancement of the relative strength of an antisymmetric pseudomagnetic field as caused by a conical intersection. In other words, exactly around a conical intersection, the nuclear dynamics is dominated by dissipation.

\section{Friction Tensor as A Function of Inverse Temperature $\beta$}\label{si:ft_T_effect}
\begin{figure}[!h]
\centering
\includegraphics[width=.55\textwidth]{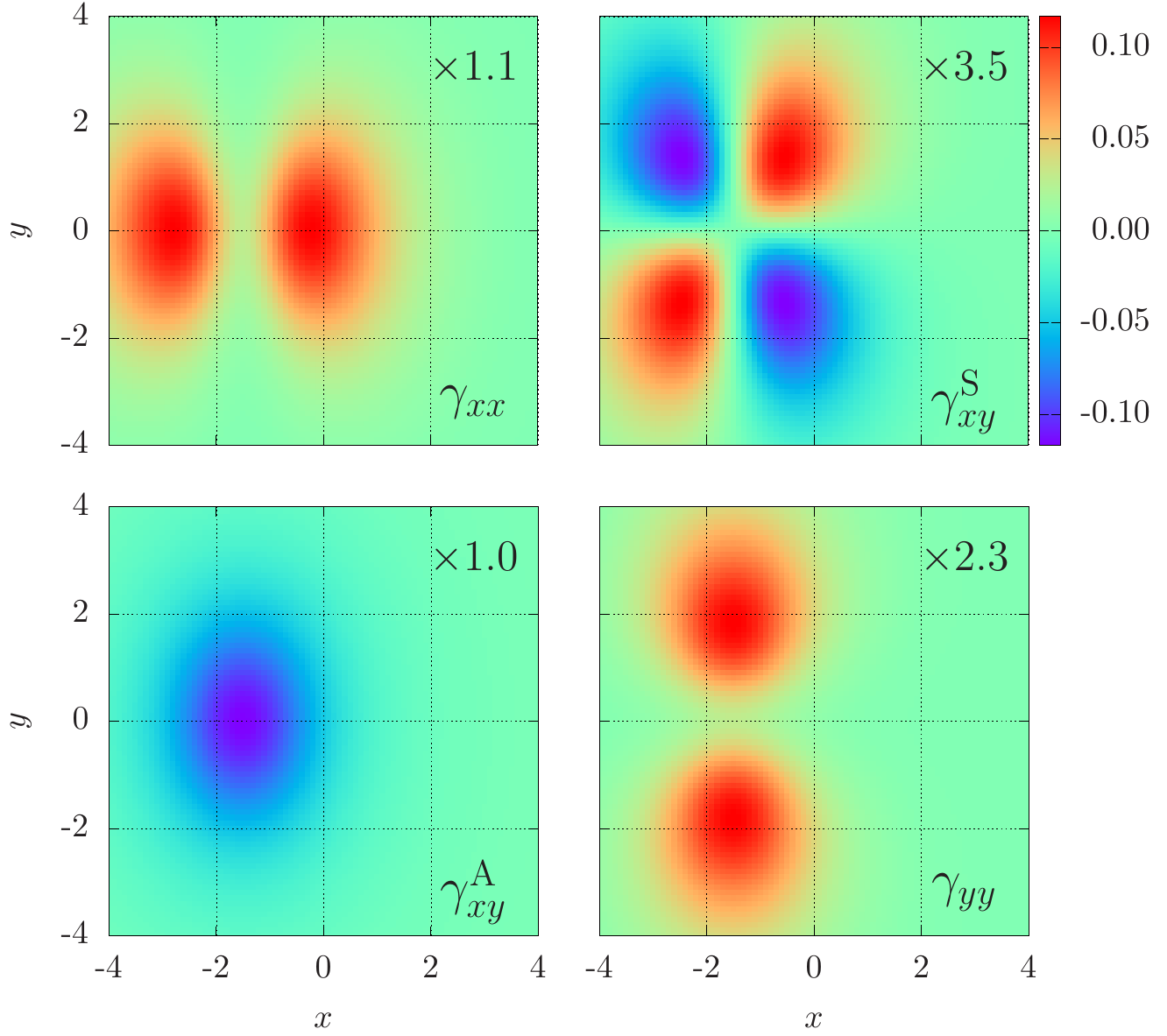}
\caption{Friction tensor calculation results: $\gamma_{xx}$ (top left), $\gamma_{xy}^{\mathrm{S}}$ (top right), $\gamma_{xy}^{\mathrm{A}}$ (bottom left) and $\gamma_{yy}$ (bottom right). Parameters: $\tilde{\Gamma}=1$, $\mu_{\mathrm{R}}=\mu_{\mathrm{L}}=0$, $\beta=1$, $A=1$, $B=1$, $\Delta=3$. Note that we have raised the temperature relative to Fig. \ref{fig:b2_D3} in the main text of the letter. We find that the relative strength of the antisymmetric friction tensor $\gamma_{xy}^{\mathrm{A}}$ has decreased, though $\gamma_{xy}^{\mathrm{A}}$ and $\gamma^{\mathrm{S}}$ are still the same order of magnitude. \label{fig:D3}}
\end{figure}

Next, we investigate the effect of temperature on the friction tensor. Note that, according to Eq. (\ref{eq:condition1_at_equil}) and (\ref{eq:condition2_at_equil}), the friction tensor in Eq. (\ref{eq:ft_tls}) can be represented as $\gamma_{\mu\nu}=\int d\epsilon\,(F_{\mu\nu}^{\mathrm{S}}+F_{\mu\nu}^{\mathrm{A}})f$. Here $F_{\mu\nu}^{\mathrm{S}}$ and $F_{\mu\nu}^{\mathrm{A}}$ are the integrands of Eqs. (\ref{eq:ft_tls_sym}) and (\ref{eq:ft_tls_antisym}) respectively (excluding the Fermi-Dirac distribution $f$), namely
\begin{alignat}{2}
F_{\mu\nu}^{\mathrm{S}}=&\frac{1}{\pi}\Big\{&&-4\epsilon\mathfrak{Re}\left\{C\tilde{\epsilon}\right\}\left(\partial_{\mu}\mathbf{h}\cdot\partial_{\nu}\mathbf{h}\right)\left(\mathbf{h}\cdot\mathbf{h}\right)\notag\\
&&&+4\epsilon\mathfrak{Re}\left\{C\tilde{\epsilon}\right\}\left(\partial_{\mu}\mathbf{h}\cdot\mathbf{h}\right)\left(\mathbf{h}\cdot\partial_{\nu}\mathbf{h}\right)\notag\\
&&&+4\epsilon\mathfrak{Re}\left\{C\tilde{\epsilon}\right\}\left(\partial_{\mu}\mathbf{h}\cdot\mathbf{h}\right)\left(\partial_{\nu}\mathbf{h}\cdot\mathbf{h}\right)\notag\\
&&&+\left(\epsilon^{2}+h^{2}+\frac{\Gamma^{2}}{4}\right)\mathfrak{Re}\left\{C\left(\tilde{\epsilon}^{2}+h^{2}\right)\right\}\partial_{\mu}\mathbf{h}\cdot\partial_{\nu}\mathbf{h}\Big\}\\
F_{\mu\nu}^{\mathrm{A}}=&\frac{1}{\pi}\Big\{&&-2\epsilon\mathfrak{Im}\left\{C\left(\tilde{\epsilon}^{2}+h^{2}\right)\right\}\partial_{\mu}\mathbf{h}\cdot\left(\partial_{\nu}\mathbf{h}\times\mathbf{h}\right)\notag\\
&&&-2\left(\epsilon^{2}+h^{2}+\frac{\Gamma^{2}}{4}\right)\mathfrak{Im}\left\{C\tilde{\epsilon}\right\}\left(\partial_{\mu}\mathbf{h}\times\mathbf{h}\right)\cdot\partial_{\nu}\mathbf{h}\Big\}
\end{alignat}
These integrands contain not only the effect of broadening from the metal (recall that $G^{<}=-i2f\mathfrak{Im}\Ropt{G}$), but also the derivatives of $h^{\mathrm{s}}$ as a function of nuclear coordinates ($\mu$, $\nu$) plus the partial derivative $\partial_{\epsilon}\Ropt{G}$. Since the temperature appears only in the Fermi-Dirac distribution, changing $\beta$ effectively controls the overlap between $F^{\mathrm{S}}$ and $f$ (and between $F^{\mathrm{A}}$ and $f$). This analysis can lead to the different orders of magnitude for the symmetric and antisymmetric friction tensors. For instance, as shown in Fig 1 of the main text, for the case $\beta=2$, the antisymmetric tensor is much larger than the symmetric tensor --- and this becomes only stronger at even lower temperatures. By comparison, in Fig. \ref{fig:D3} we show the contour plots at a higher temperature $\beta=1$, where the antisymmetric friction tensor is now of the same order of magnitude as the symmetric friction tensor.  For a plot of one numerical example of $F(\epsilon)$, see Figs. \ref{fig:eps_integrands_F_n_f_diff_T} below.

As a side note, beyond temperature effects, we mention that one could also utilize the chemical potential $\mu$ to control $\gamma^{\mathrm{S}}$ and $\gamma^{\mathrm{A}}$. For instance, since $F^{\mathrm{S}}$ and $F^{\mathrm{A}}$ are both odd functions of $\epsilon$, when the chemical potential is high or low enough, both $\gamma^{\mathrm{S}}$ and $\gamma^{\mathrm{A}}$ disappear.

\section{Friction Tensor as A Function of Metal-Molecule Coupling $\tilde{\Gamma}$}\label{si:ft_other_results}



\begin{figure}[h!]
\centering
\includegraphics[width=.55\textwidth]{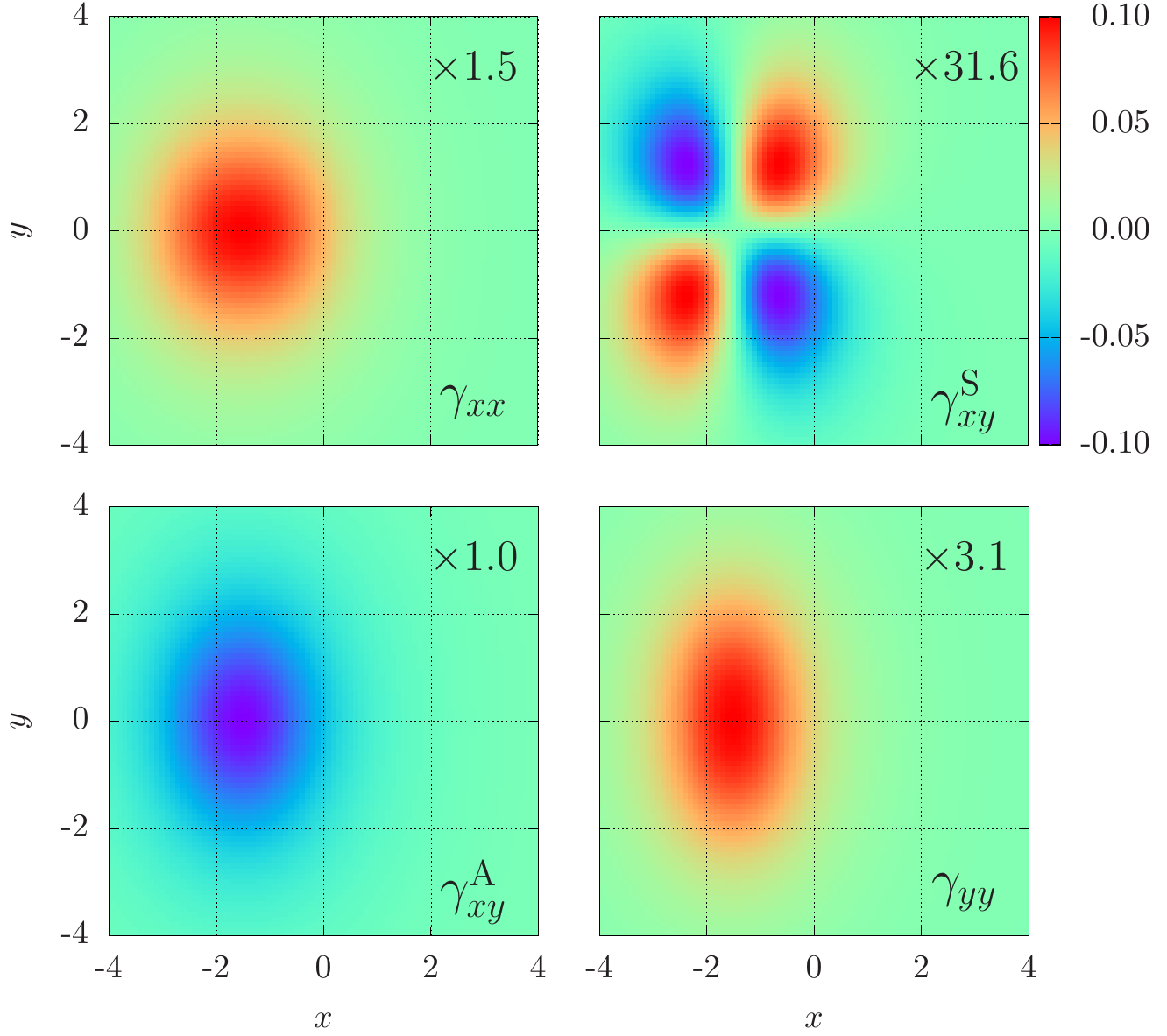}
\caption{Friction tensor calculation results as a comparison of different $\tilde{\Gamma}$'s: $\gamma_{xx}$ (top left), $\gamma_{xy}^{\mathrm{S}}$ (top right), $\gamma_{xy}^{\mathrm{A}}$ (bottom left) and $\gamma_{yy}$ (bottom right). Parameters: $\tilde{\Gamma}=3$, $\mu_{\mathrm{R}}=\mu_{\mathrm{L}}=0$, $\beta=2$, $A=1$, $B=1$, $\Delta=3$. Note that, as $\tilde{\Gamma}$ increases relative to Fig. \ref{fig:b2_D3} in the main text of the letter, the relative strength of the antisymmetric friction tensor $\gamma_{xy}^{\mathrm{A}}$ is effectively unchanged.\label{fig:G1_D1}}
\end{figure}

\begin{figure}[h!]
\centering
\includegraphics[width=.55\textwidth]{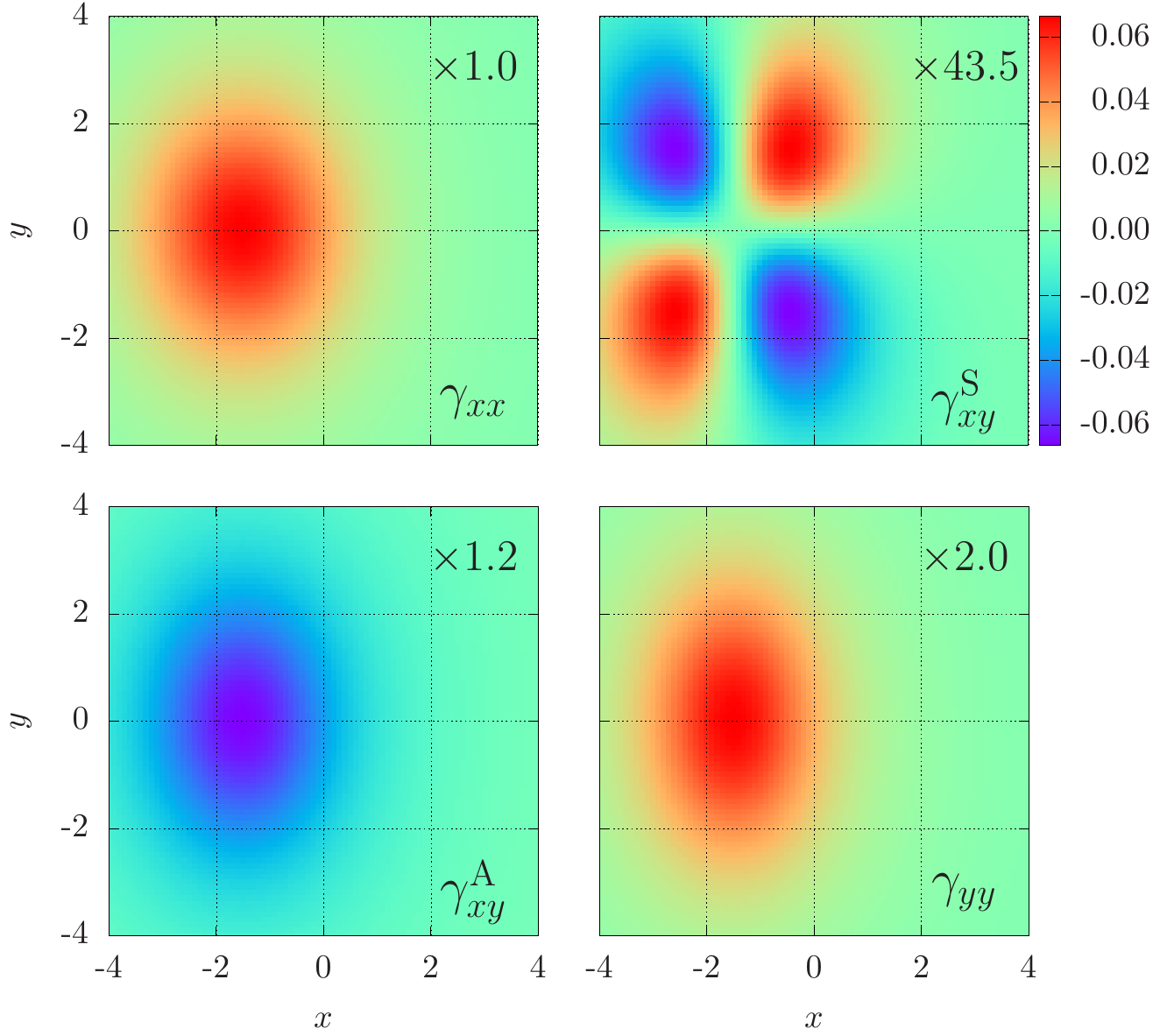}
\caption{Friction tensor calculation results  as a comparison of different  $\tilde{\Gamma}$'s: $\gamma_{xx}$ (top left), $\gamma_{xy}^{\mathrm{S}}$ (top right), $\gamma_{xy}^{\mathrm{A}}$ (bottom left) and $\gamma_{yy}$ (bottom right). Parameters: $\tilde{\Gamma}=5$, $\mu_{\mathrm{R}}=\mu_{\mathrm{L}}=0$, $\beta=2$, $A=1$, $B=1$, $\Delta=3$. Note that when $\tilde{\Gamma}$ gets very large, both the antisymmetric and symmetric
components of the friction tensor decrease in magnitude.\label{fig:G2_D1}}
\end{figure}

For molecules in the gas phase, it is fairly standard to ascertain the size of a Berry force from
the relevant derivative couplings\cite{berry1993chaotic}. In this manuscript, however, our goal has been
to report the size of the Berry force in a condensed environment. Thus, 
for completeness, in Figs. \ref{fig:G1_D1} and \ref{fig:G2_D1}, we plot the friction tensor as a function of 
$\tilde{\Gamma}$, which represents coupling of the molecule to the metal. In Fig. \ref{fig:G1_D1}, we increase $\tilde{\Gamma}$ relative to Fig. \ref{fig:b2_D3} in the main text of the letter. We find that the relative strength of the antisymmetric friction tensor $\gamma_{xy}^{\mathrm{A}}$ is effectively unchanged. 
Furthermore, in general,
when $\tilde{\Gamma}$ gets very large, both the antisymmetric and symmetric
components of the friction tensor decrease, as shown in Fig. \ref{fig:G2_D1}.

\section{One Proposed Experimental Realization: A Diphenylmethane Junction}\label{si:diphenylmethane}
\begin{figure}[!h]
\centering
\includegraphics[width=.35\textwidth]{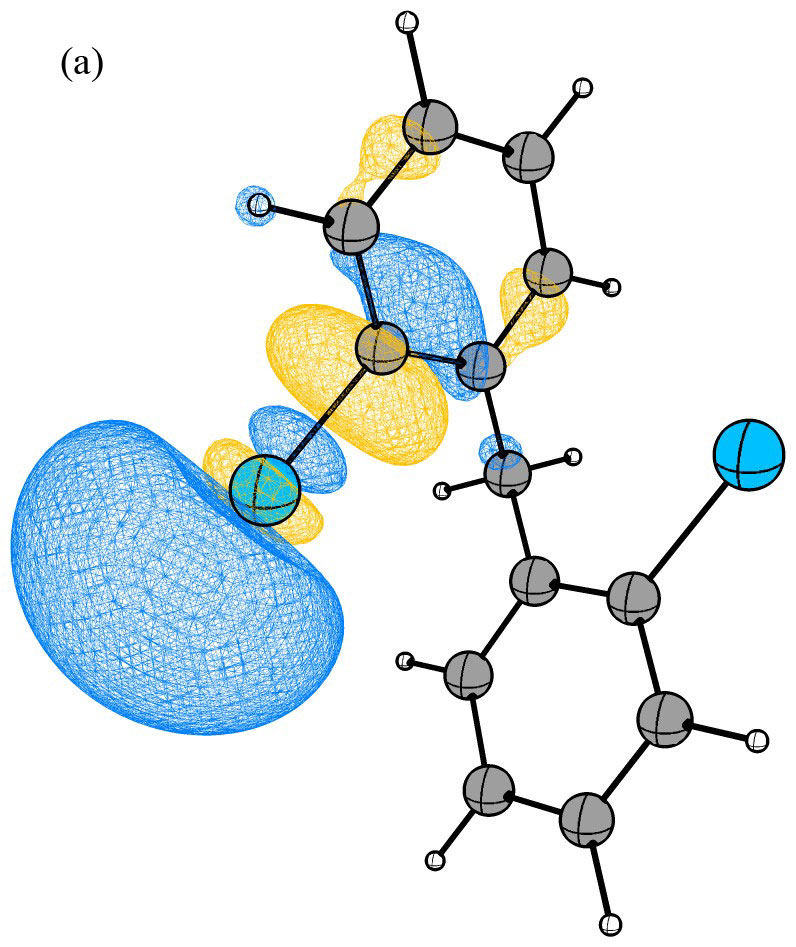}\qquad
\includegraphics[width=.35\textwidth]{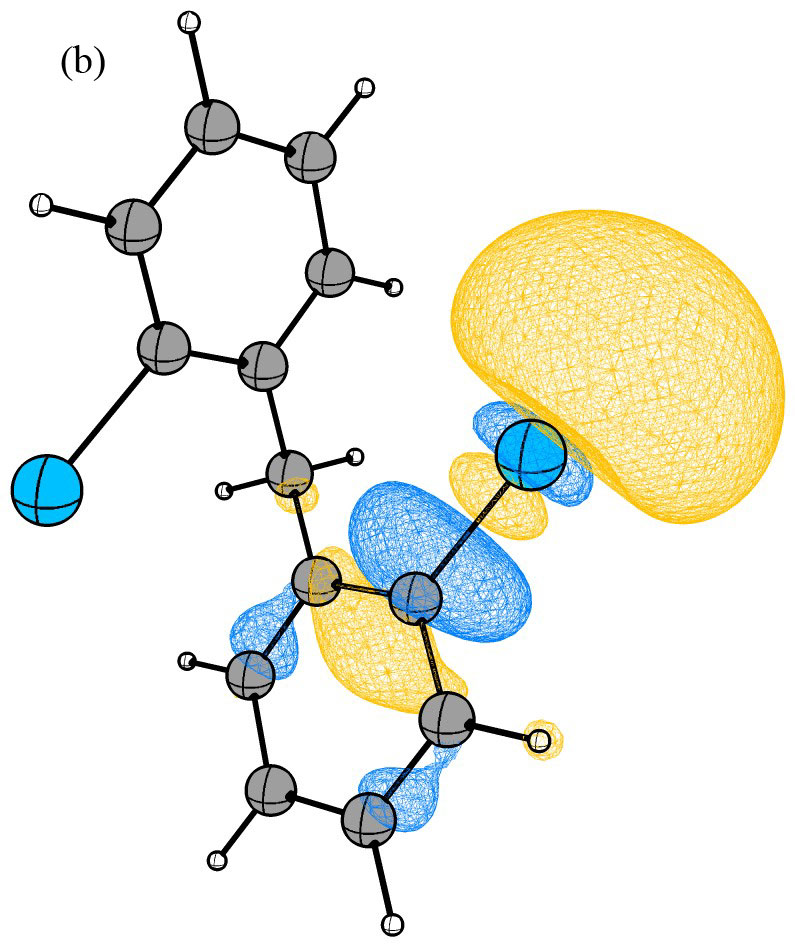}
\caption{Optimized geometry of diphenylmethane with two hydrogen atoms (white) replaced by two gold atoms (blue), which heuristically represent two gold clusters. We map the two localized orbitals in (a) and (b) to the two diabats used to construct the system Hamiltonian $\mathbf{h}^{\mathrm{s}}$ in the main body of the text.\label{fig:opt_geo_bpm2au_diab_orb}}
\end{figure}

In this section, we would like to show how the theory above can be applied within a realistic calculation of a molecule.  We focus on the diphenylmethane (which can be easily prepared by a Friedel–Crafts reaction\cite{yin2006synthesis}). We imagine a scenario whereby the molecule is placed in a junction between gold electrodes, and for our \textit{ab initio} simulations below, we replace two hydrogen atoms  by two gold atoms.  As shown in Fig. \ref{fig:opt_geo_bpm2au_diab_orb} (a), the optimized geometry has nearly a $C_{2}$ symmetry. We perform DFT calculations with the B3LYP exchange functional, LANL2DZ basis and LANL2DZ effective core potential. Among the orbital energies, the LUMO ($-0.0874\,\mathrm{a.u.}$) and LUMO+1 ($-0.0853\,\mathrm{a.u.}$) are well separated from all other orbitals (HOMO: $-0.2333\,\mathrm{a.u.}$, LUMO+2: $-0.0222\,\mathrm{a.u.}$), which allows us to consider the single-particle Hamiltonian with explicitly two orbitals as described in the main body of the text.  Experimentally these two orbitals can be probed by tuning the chemical potential of the nearby electrodes.

\begin{figure}[!h]
\centering
\includegraphics[width=.3\textwidth]{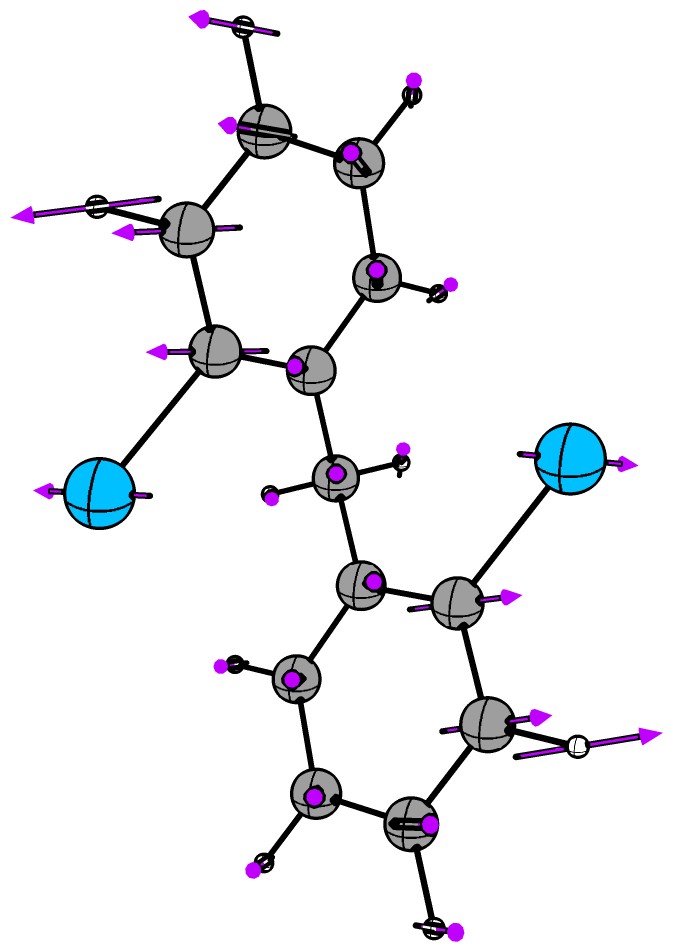}
\caption{Normal mode of the diphenylmethane with the lowest frequency.\label{fig:normal_mode_lowest_freq}}
\end{figure}

\begin{figure}[!h]
\centering
\includegraphics[width=.9\textwidth]{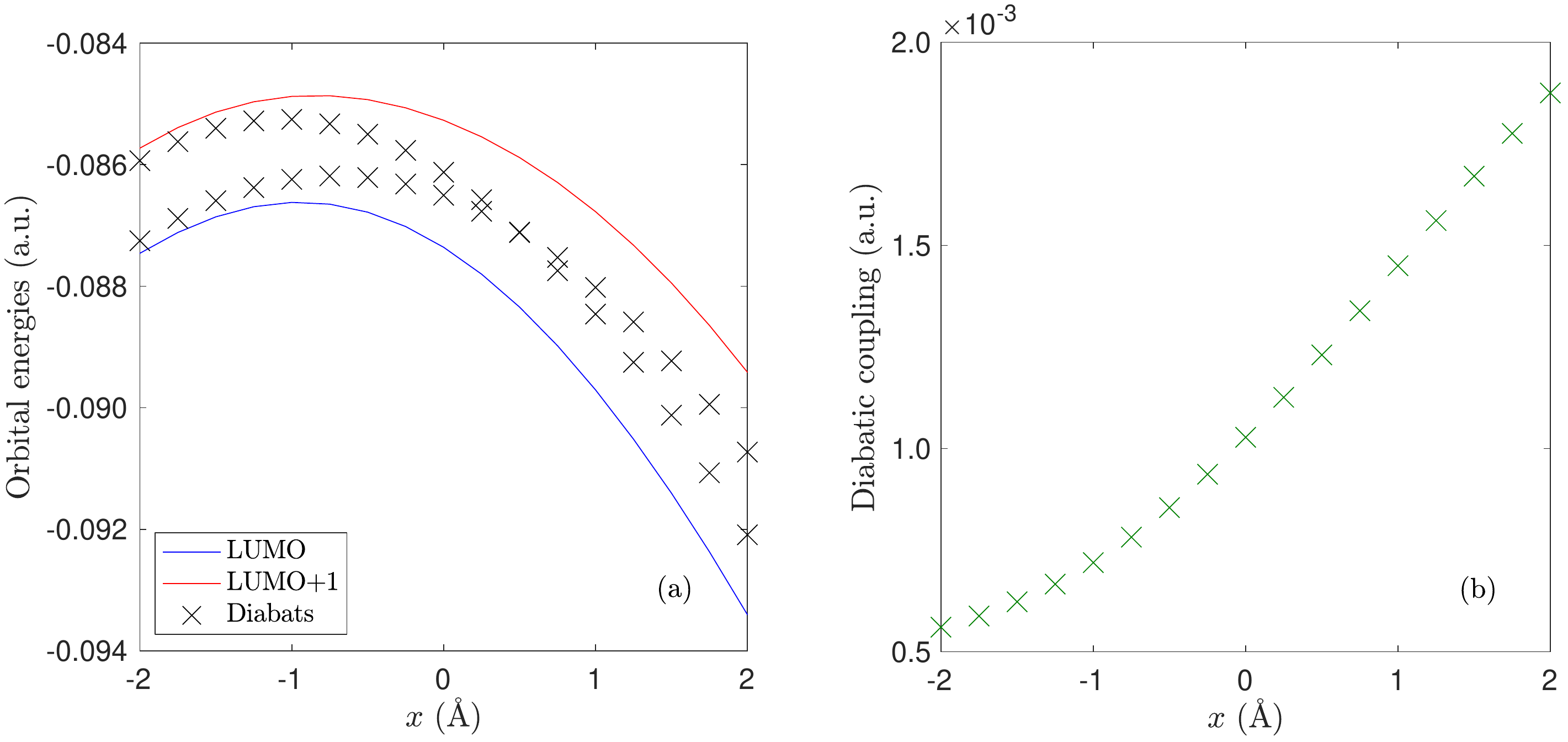}
\caption{(a) Orbital energies of LUMO (blue) and LUMO+1 (red) along the lowest normal mode. The two diabats (black cross) are obtained through standard Boys localization, and (b) is the corresponding diabatic coupling.\label{fig:adiabats_diabats_diabatic_coupling}}
\end{figure}

Next, in order to apply the friction model constructed in the main body of the text, we further apply a  normal mode analysis to isolate the low frequency modes of the molecule. Obviously, diphenylmethane has $3\times25-6=69$ normal modes, and the friction tensor is a large matrix. That being said, our goal here is to show that the antisymmetric component can be crucial, and to that end, it will suffice to focus on only a handful of normal modes that are thermally populated at reasonable temperatures. 
We will consider the normal mode of lowest frequency plotted in Fig. \ref{fig:normal_mode_lowest_freq} which has energy $17.3\,\mathrm{cm}^{-1}$.  Over this mode, we scan the orbital energies of the LUMO and LUMO+1, and follow the standard Boys localization procedure \cite{subotnik2008constructing} to obtain two diabats and a corresponding diabatic coupling. In Fig. \ref{fig:adiabats_diabats_diabatic_coupling}, we plot the orbital energies of LUMO (blue), LUMO+1 (red), the two localized diabat energies (black cross)  and the corresponding diabatic coupling; the diabatic orbital information will be used in constructing the system Hamiltonian $\mathbf{h}^{\mathrm{s}}$ in the main body of the text.  Notice that, as shown in Fig. \ref{fig:opt_geo_bpm2au_diab_orb}, the two diabats $1$ and $2$ are localized near the two different gold atoms, which justifies the assumption that the hybridization function ($\Gamma_{mn}$) be diagonal;  $\Gamma_{12} = 0$.

\begin{figure}[!h]
\centering
\includegraphics[width=.4\textwidth]{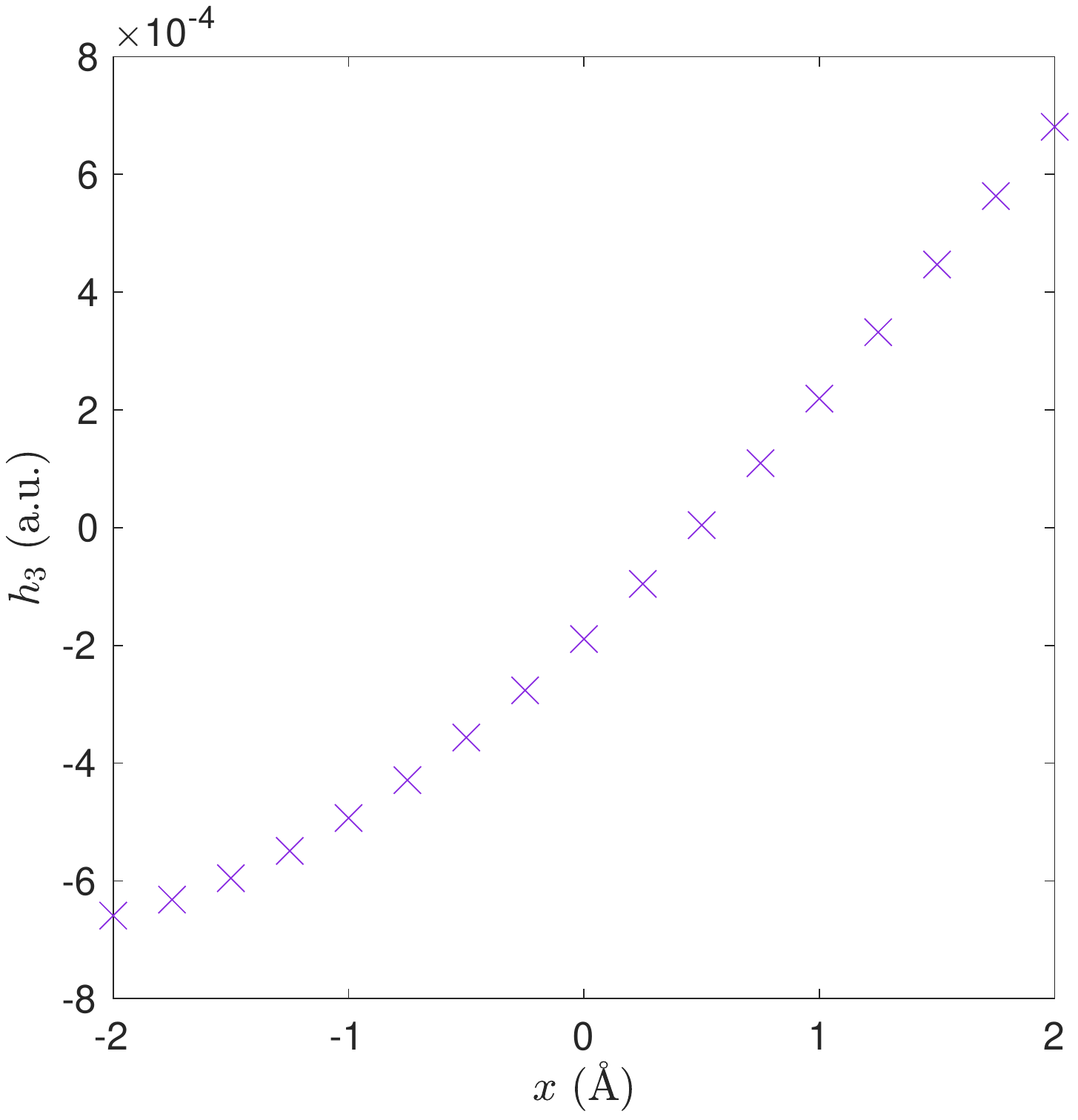}
\caption{The $\sigma_{3}$ component, $h_{3}$, used for the friction tensor calculations.\label{fig:h3}}
\end{figure}

The two diabats and the diabatic coupling in Fig. \ref{fig:adiabats_diabats_diabatic_coupling} form the system Hamiltonian $\mathbf{h}^{\mathrm{s}}$ in the main body of the text. The diagonal terms of $\mathbf{h}^{\mathrm{s}}$ can be divided into symmetric and antisymmetric parts, $h_{0}I$ and $h_{3}\sigma_{3}$ respectively, and we disregard the symmetric part since it does not affect the friction tensor. The antisymmetric term is shown in Fig. \ref{fig:h3}. As one can observe in Fig. \ref{fig:adiabats_diabats_diabatic_coupling} (b) and Fig. \ref{fig:h3}, the diabatic coupling and $h_{3}$ are not exactly linear functions. Nevertheless, for simplicity and in order to be consistent with the main text, we will make a linear approximation so as to estimate the order of magnitude of the parameters in our model. The linear functions for $h_{3}$ and the diabatic coupling are fit to $\lambda x+\Delta$ and $Ax+C$ respectively where $\lambda=3.44\times10^{-4}$, $\Delta=-1.13\times10^{-4}$, $A=3.44\times10^{-4}$ and $C=1.11\times10^{-3}$ (all in the a.u.). Note that the room temperature is $k_{\mathrm{B}}T=25.7\,\mathrm{meV}=9.45\times10^{-4}\,\mathrm{a.u.}$, and low temperature experiments can be easily done nowadays.
We will estimate $\tilde{\Gamma}$ according to Refs. \citenum{guo2012spin,koseki2019spin,varela2016effective}, assuming the value ranges from $1-100\,\mathrm{meV}$. (In Ref. \citenum{guo2012spin}, the authors set the spin-orbit interaction to be one order smaller than the diabatic coupling).

At this point, the only unknown in our model is the strength of the spin-orbit coupling.  Unfortunately, estimating matrix elements rigorously is very demanding and convergence is not easy to achieve: are fully relativistic two-component simulations necessary (or can we use perturbation theory on top of spin-restricted spin-collinear DFT)?  Gold atoms introduce a great deal of spin-orbit coupling, but how many gold atoms must be included in a cluster for convergence?  Answering these questions remains a hot topic within the field\cite{doi:10.1021/acs.jctc.0c00621}. Note also that calculations become much more demanding with increasing system size, such that implementing a high level code make sampling nuclear configurations quite expensive.  For all of these reasons, we will simply 
follow Refs. \citenum{guo2012spin,koseki2019spin,varela2016effective} and fit the spin-orbit interaction to be of the form by $By+C'$ where the parameters are chosen as as either the same order of magnitude as the diabatic coupling ($B=A$, $C'=C$) or  $5\%$ times smaller than the diabatic coupling ($B=0.05A$ and $C'=0.05C$) (which is smaller than any estimates in Refs. \citenum{guo2012spin,koseki2019spin,varela2016effective}).

\begin{figure}[!h]
\centering
\includegraphics[width=.55\textwidth]{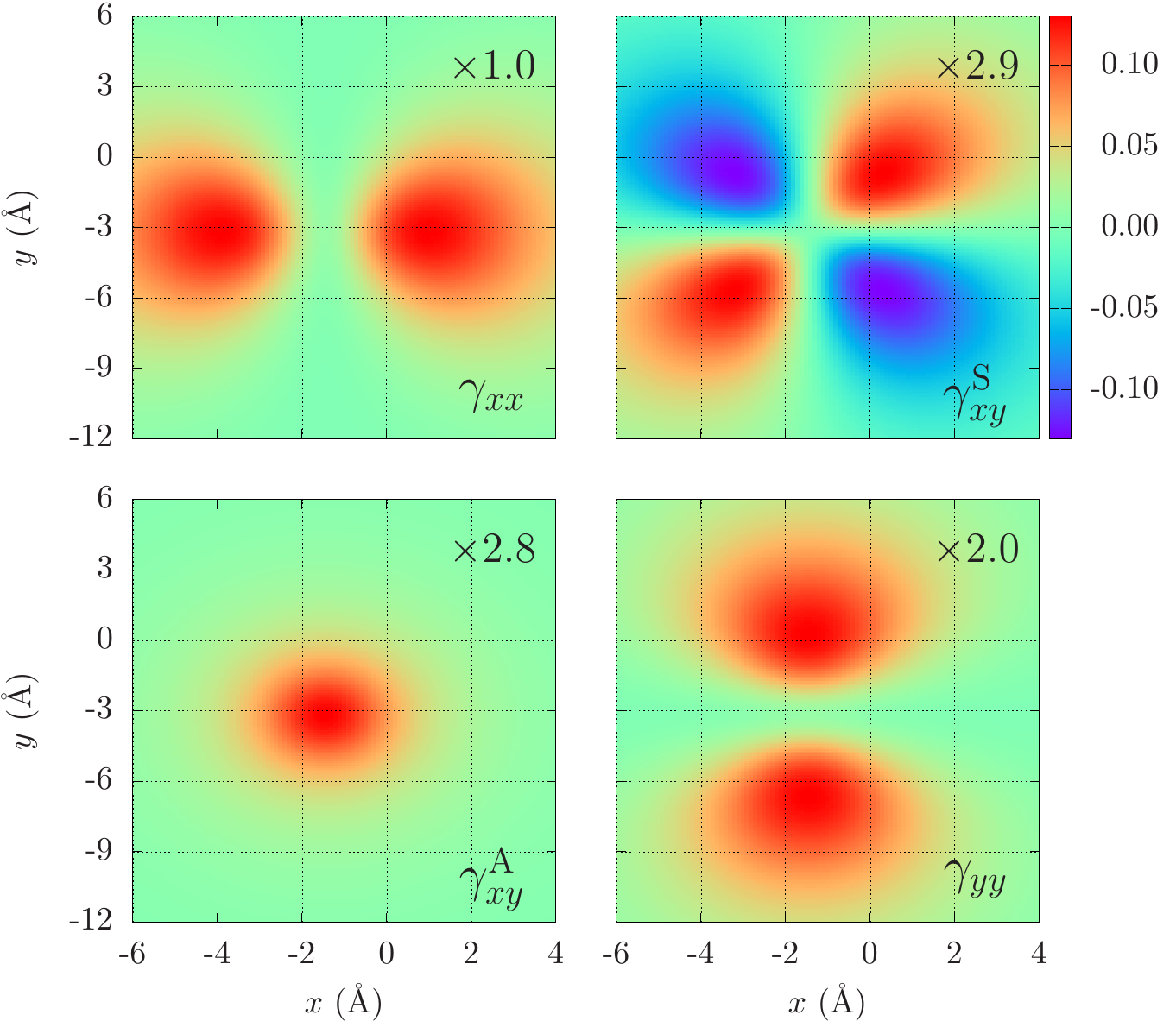}
\caption{Friction tensor calculation results: $\gamma_{xx}$ (top left), $\gamma_{xy}^{\mathrm{S}}$ (top right), $\gamma_{xy}^{\mathrm{A}}$ (bottom left) and $\gamma_{yy}$ (bottom right). Parameters: $\tilde{\Gamma}=10\,\mathrm{meV}$ and $B=A$ under room temperature.\label{fig:BA_G1_roomT}}
\end{figure}

In the following, we provide some representative results. In Fig. \ref{fig:BA_G1_roomT}, we show contour plots for the friction tensor with $\tilde{\Gamma}=10\,\mathrm{meV}$ and $B=A$ under room temperature. The antisymmetric friction tensor is comparable to the symmetric friction tensors and this phenomenon is still robust for larger $\tilde{\Gamma}=100\,\mathrm{meV}$. Also notice that the effective regions for the two kinds of friction tensors are quite different so that the nuclear motion will be strongly affected by the pseudo magnetic field. 

\begin{figure}[!h]
\centering
\includegraphics[width=.55\textwidth]{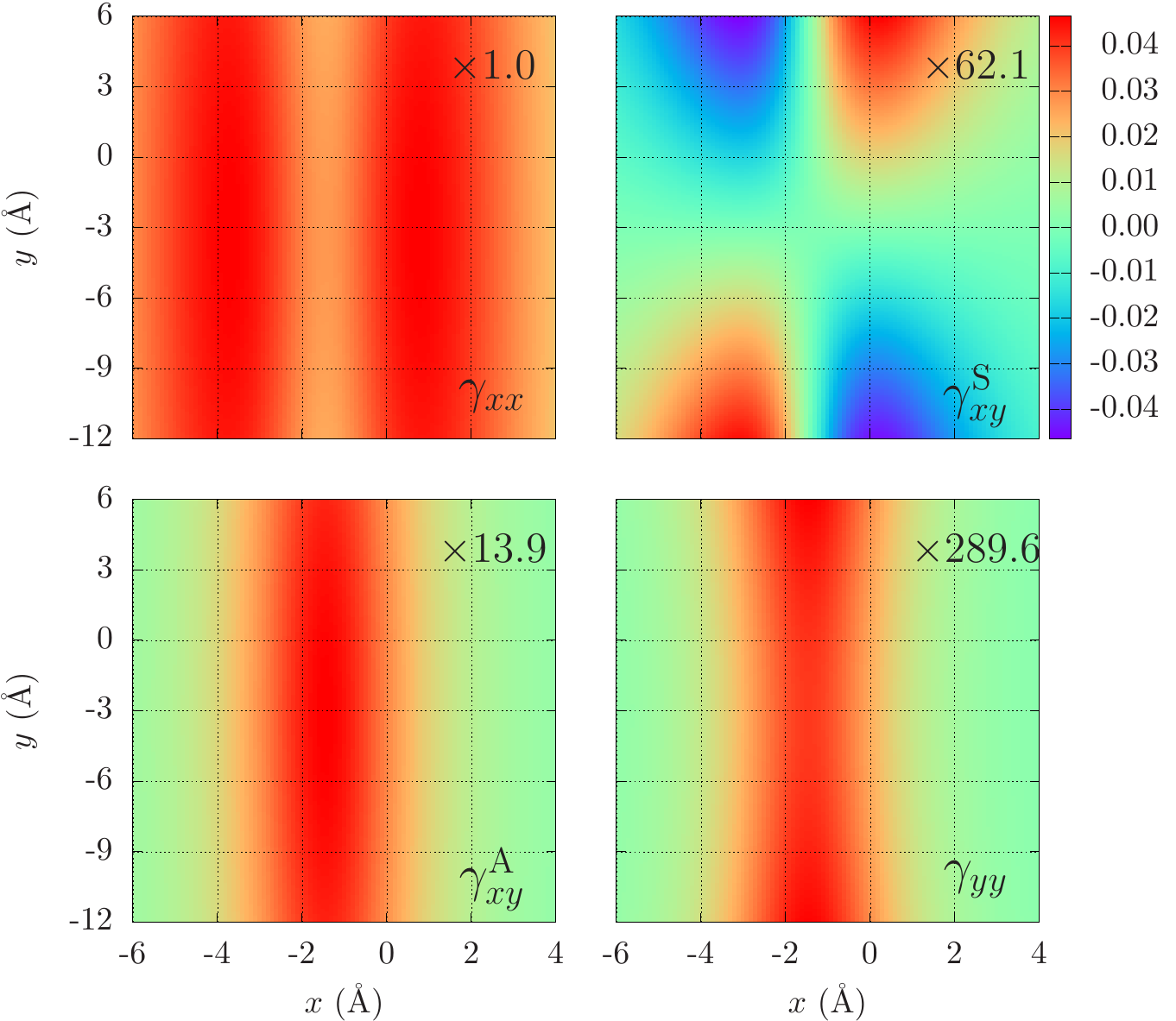}
\caption{Friction tensor calculation results: $\gamma_{xx}$ (top left), $\gamma_{xy}^{\mathrm{S}}$ (top right), $\gamma_{xy}^{\mathrm{A}}$ (bottom left) and $\gamma_{yy}$ (bottom right). Parameters: $\tilde{\Gamma}=30\,\mathrm{meV}$ and $B=0.1A$ under room temperature.\label{fig:B01A_G3_roomT}}
\end{figure}

\begin{figure}[!h]
\centering
\includegraphics[width=.55\textwidth]{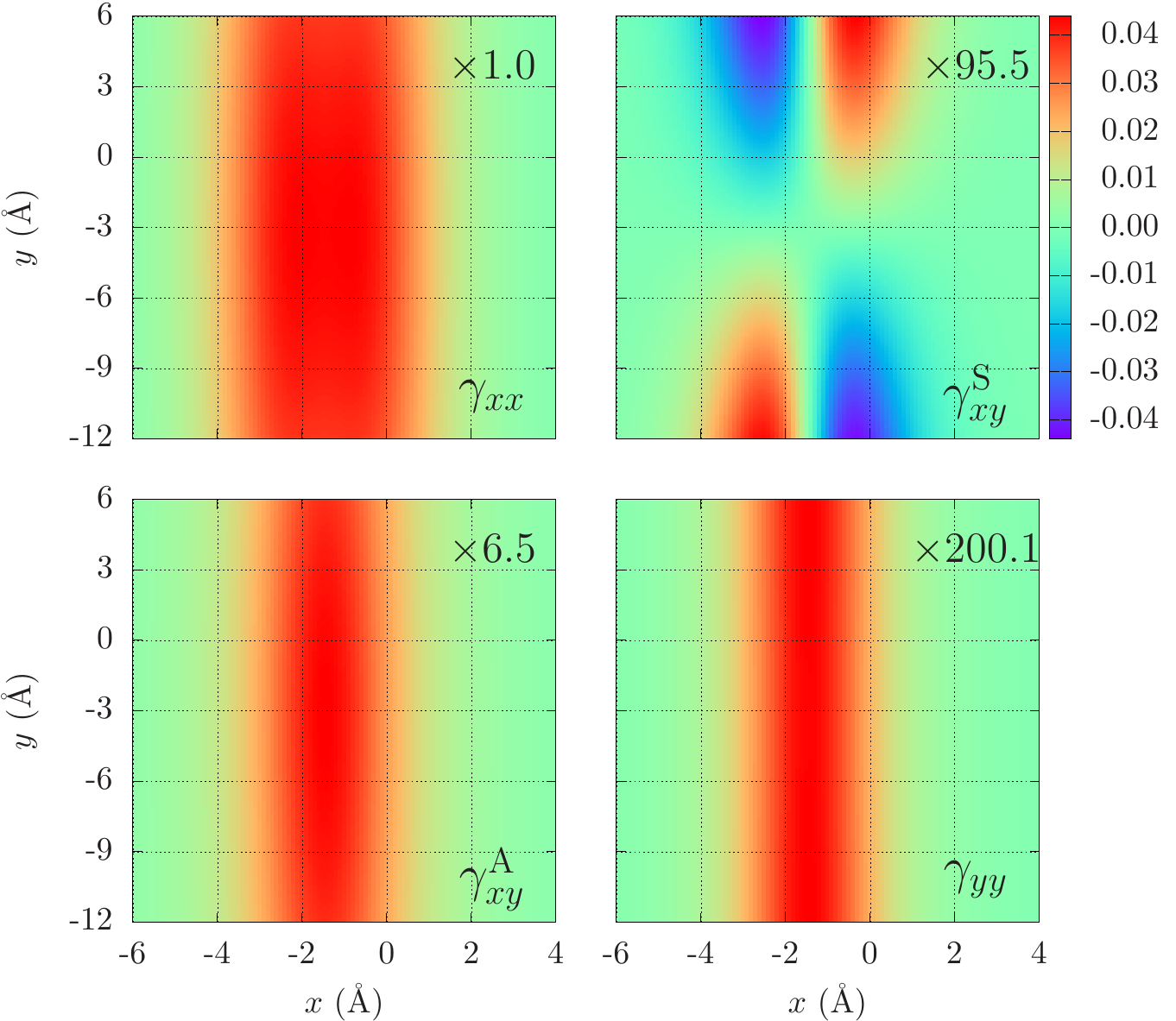}
\caption{Friction tensor calculation results: $\gamma_{xx}$ (top left), $\gamma_{xy}^{\mathrm{S}}$ (top right), $\gamma_{xy}^{\mathrm{A}}$ (bottom left) and $\gamma_{yy}$ (bottom right). Parameters: $\tilde{\Gamma}=30\,\mathrm{meV}$, $k_{\mathrm{B}}T=8.6\,\mathrm{meV}$ and $B=0.1A$.\label{fig:B01A_G3_Tdb3}} 
\end{figure}

\begin{figure}[!h]
\centering
\includegraphics[width=.55\textwidth]{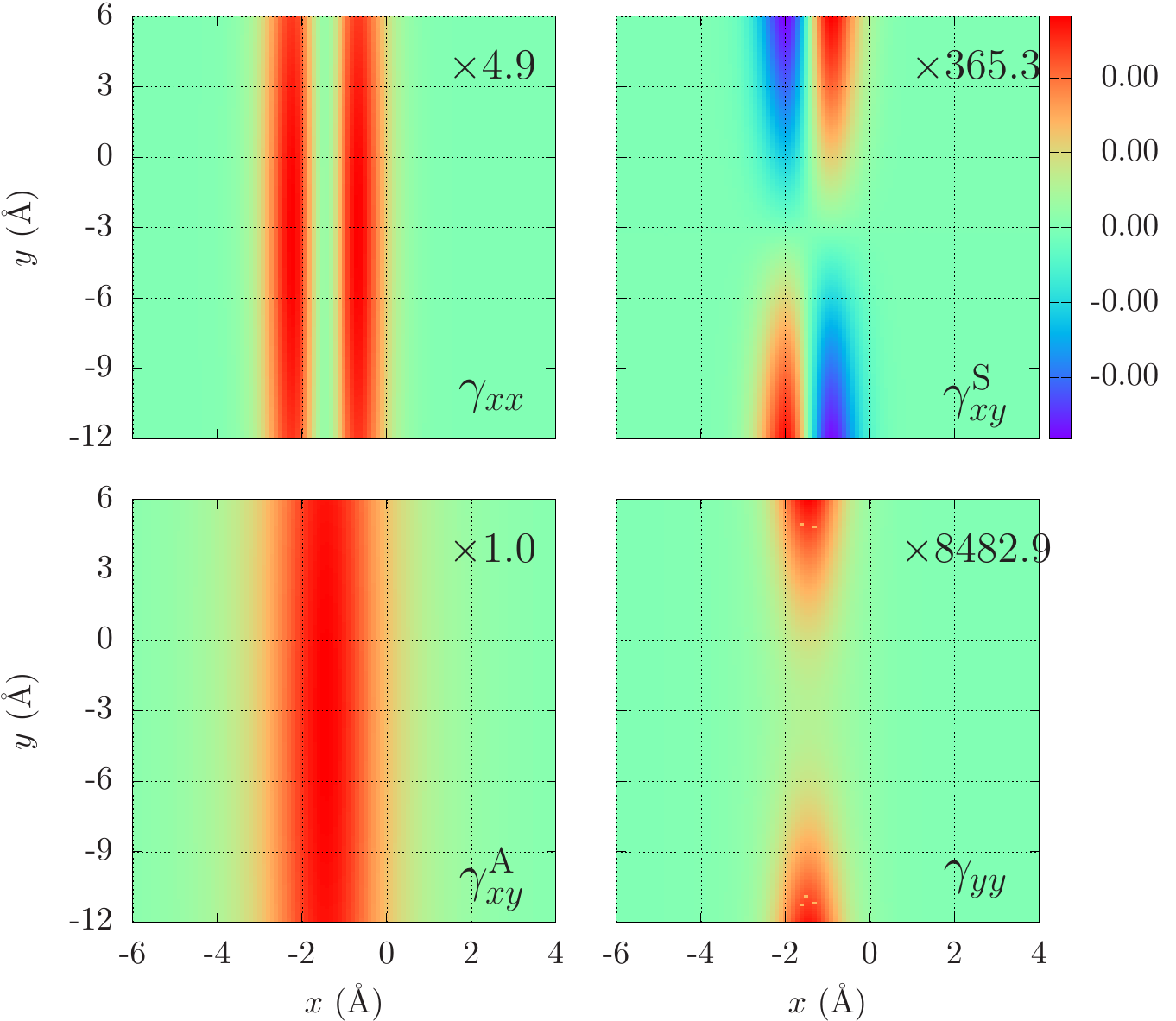}
\caption{Friction tensor calculation results: $\gamma_{xx}$ (top left), $\gamma_{xy}^{\mathrm{S}}$ (top right), $\gamma_{xy}^{\mathrm{A}}$ (bottom left) and $\gamma_{yy}$ (bottom right). Parameters: $\tilde{\Gamma}=1\,\mathrm{meV}$, $k_{\mathrm{B}}T=2.6\,\mathrm{meV}$ and $B=0.05A$.\label{fig:B005A_G3_Tdb10}} 
\end{figure}

\begin{figure}[!h]
\centering
\includegraphics[width=.95\textwidth]{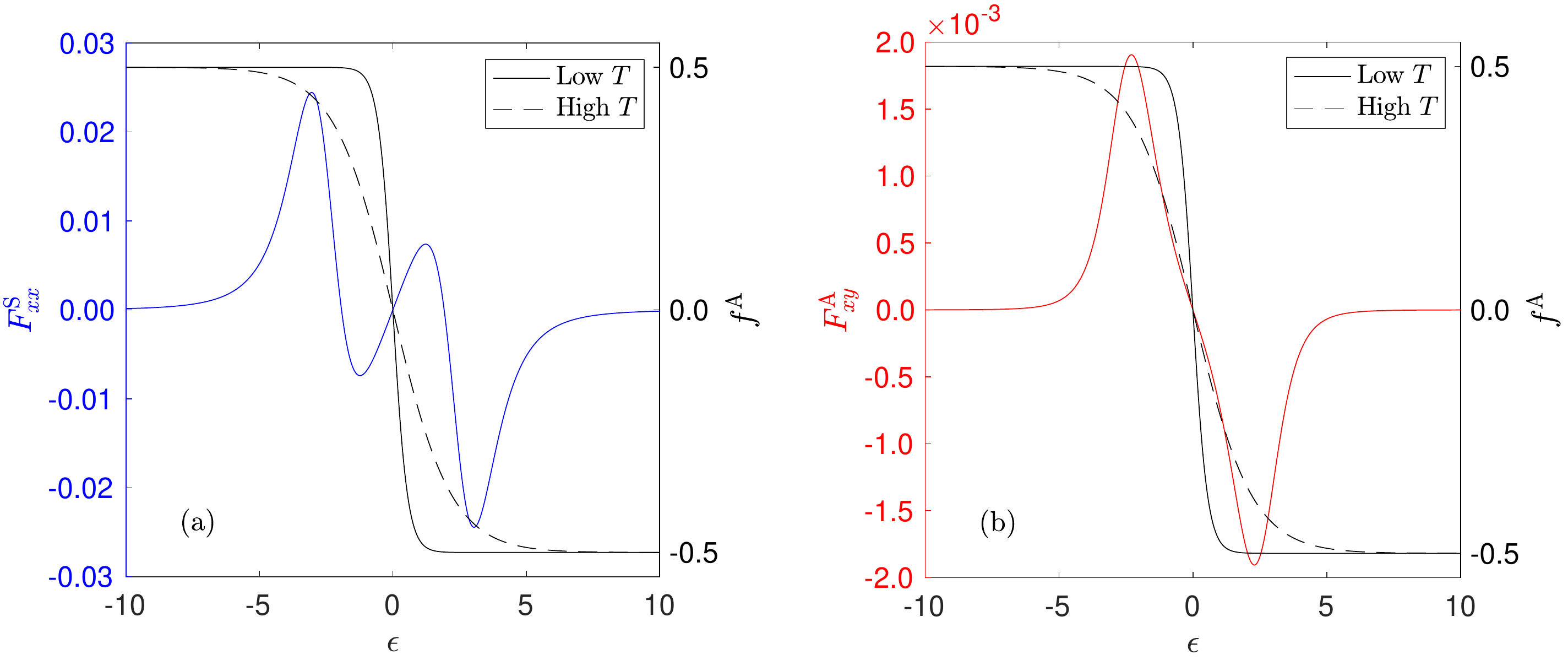}
\caption{Integrands for the diphenylmethane at one geometry. Blue and red lines represent the integrands $F_{xx}^{\mathrm{S}}$ and $F_{xy}^{\mathrm{A}}$ respectively as defined in Sec. \ref{si:ft_T_effect}. Black lines (solid for low $T$, dash for high $T$) represent the antisymmetrized Fermi distributions $f^{\mathrm{A}}$. As the temperature is lower, the symmetric friction tensor becomes smaller and the antisymmetric friction tensor becomes larger. Parameters: $x=-1.4$, $y=-3.2$, $\tilde{\Gamma}=30\,\mathrm{meV}$, $B=0.05A$.\label{fig:eps_integrands_F_n_f_diff_T}}
\end{figure}

Next, we address the common situation whereby the spin-orbit coupling matrix elements are small (i.e. much smaller than the diabatic couplings).  In Fig. \ref{fig:B01A_G3_roomT}, we show contour plots for the friction tensor with $B=0.1A$  (still under room temperature and $\tilde{\Gamma}=30\,\mathrm{meV}$). Observe that the antisymmetric friction tensor becomes one order smaller than the symmetric ones. Nevertheless, when the temperature is decreased, $\gamma_{xy}^{\mathrm{A}}$ gets relatively larger as shown in Fig. \ref{fig:B01A_G3_Tdb3}. In fact, for this case, changing temperature can make $\gamma_{xy}^{\mathrm{A}}$ dominant even when the spin-orbit interaction is extremely small ($B=0.05A$) (especially if we reduce $\tilde{\Gamma}$) as shown in Fig. \ref{fig:B005A_G3_Tdb10}. In Fig. \ref{fig:eps_integrands_F_n_f_diff_T}, in order to further explore the effect of temperature on this system,  we plot the integrands $F_{xx}^{\mathrm{S}}$ and $F_{xy}^{\mathrm{A}}$ (as defined in Sec. \ref{si:ft_T_effect}) for one specific geometry. As mentioned in Sec. \ref{si:ft_T_effect}, both $F_{xx}^{\mathrm{S}}$ and $F_{xy}^{\mathrm{A}}$  are antisymmetric functions, so that when we integrate $\int d\epsilon\,(F_{\mu\nu}^{\mathrm{S}}+F_{\mu\nu}^{\mathrm{A}})f$ so as to evaluate the friction tensor, we require only the antisymmetric part of the Fermi distribution, $f^{\mathrm{A}}$, as shown in Fig. \ref{fig:eps_integrands_F_n_f_diff_T}. For this one geometry, it is clear that, as the temperature is lowered, the symmetric friction tensor becomes smaller due to a cancellation between positive and negative integral areas, and the antisymmetric friction tensor becomes larger. The generality of these findings (and surprising temperature effects on friction) will need to be addressed and investigated in more detail for many more molecules and geometries in the future.

\begin{figure}[!h]
\centering
\includegraphics[width=.45\textwidth]{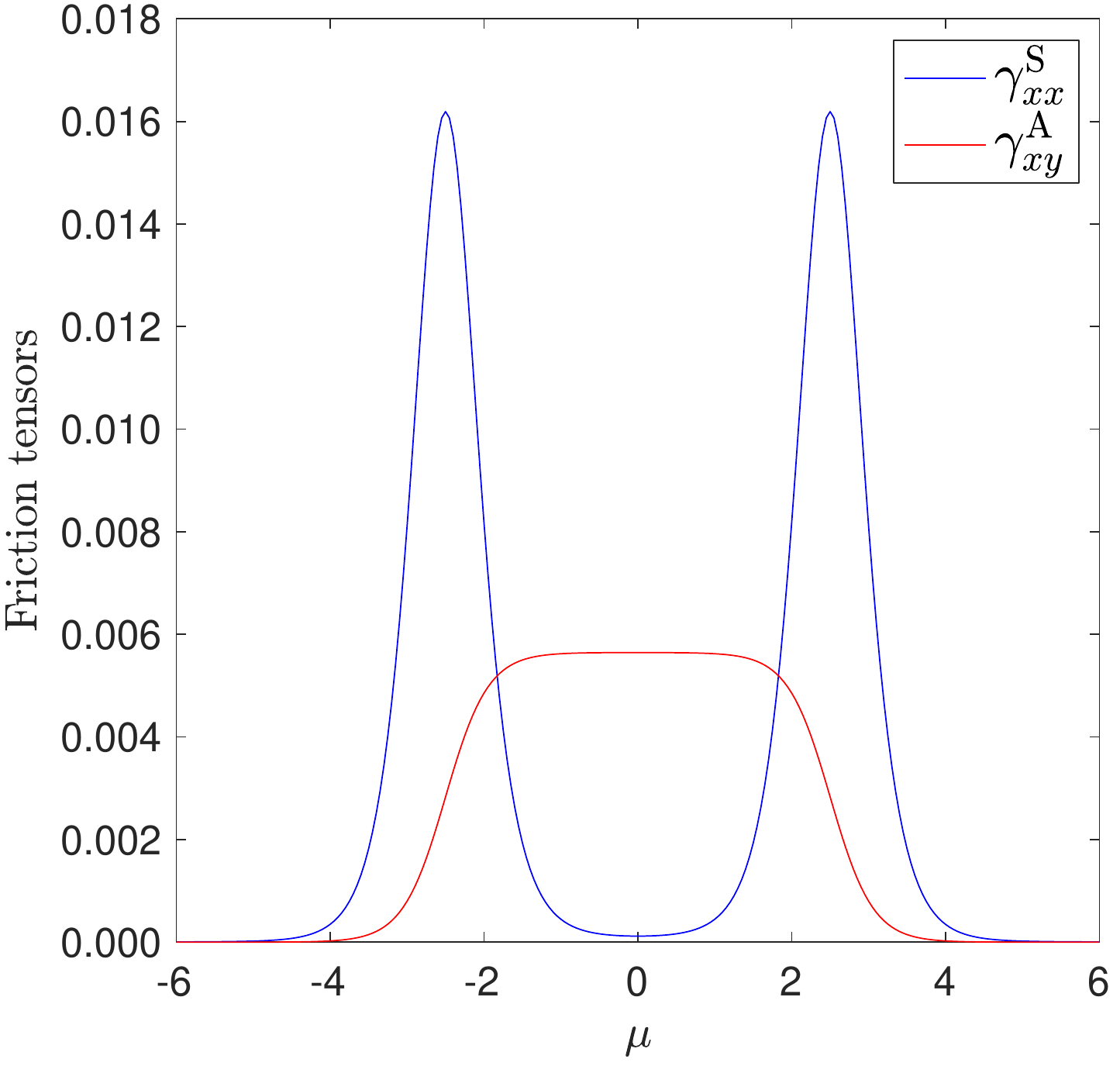}
\caption{Friction tensors as a function of $\mu$ for the diphenylmethane at one geometry. Blue and red lines represent $\gamma_{xx}^{\mathrm{S}}$ and $\gamma_{xy}^{\mathrm{A}}$ respectively. As the chemical potential is away from zero, the symmetric friction tensor becomes larger and the antisymmetric friction tensor becomes smaller. Both friction tensors approach zero when $\mu\rightarrow\pm\infty$. Parameters: $x=-1.4$, $y=-3.2$, $\tilde{\Gamma}=1\,\mathrm{meV}$, $k_{\mathrm{B}}T=2.6\,\mathrm{meV}$, $B=0.05A$.\label{fig:mu_gammaS_n_gammaA}}
\end{figure}

Lastly, we investigate how the chemical potential $\mu_{L}=\mu_{R}=\mu$ affects the friction tensors. In Fig. \ref{fig:mu_gammaS_n_gammaA}, we plot both symmetric and antisymmetric friction tensors as a function of the chemical potential $\mu$. When $\mu$ is away from zero,  the symmetric friction tensor becomes larger and the antisymmetric friction tensor becomes smaller. Both friction tensors approach zero when $\mu\rightarrow\pm\infty$, which has been discussed in SM \ref{si:ft_T_effect}.

In summary, we have demonstrated that the antisymmetric friction tensor $\gamma_{xy}^{\mathrm{A}}$ can be as large and sometimes larger than the 
symmetric friction tensor for a real system, even when the spin-orbit interaction is small (at least at low temperatures).  Therefore, for the case of a molecule on a metal surface, one must expect that the contribution of this tensor component cannot be ignored when we consider the rate of vibrational relaxation. As far as understanding the implications for a molecule between two metals (as in the diphenylmethane example here), the possibility that such an antisymmetric force will generate a spin-current is also an intriguing possibility, one that will be investigated in a future publication as well. 


\end{document}